\UseRawInputEncoding
\documentclass[3p,times]{elsarticle}
 
 \usepackage{etex}
\usepackage{lineno,hyperref} 
\modulolinenumbers[5] 
 
\usepackage{mathtools}
\usepackage{comment}
\usepackage[usenames, dvipsnames]{xcolor}
\usepackage{subfig}

\newcommand{\etal}{{\it et al.}}
\usepackage{tikz}
\usepackage{tikz-qtree}
\usetikzlibrary{arrows,decorations.markings,backgrounds,positioning,fit,shapes,patterns,fadings,calc,shadows.blur}
\usepackage{tikz-3dplot}
\usepackage{mathabx}

\usepackage{sansmath}
\usepackage{bm}
\usetikzlibrary{shadings,intersections,fadings}
\usetikzlibrary{shapes.geometric, arrows.meta}
\usetikzlibrary{calc}
\tikzset{
mystyle/.style={
  text width=3cm,
  }
}
\tikzset{box/.style={draw, rectangle, rounded corners, thick, node distance=7em, text width=6em, text centered, minimum height=3.5em}}
\tikzset{container/.style={draw, rectangle, dashed, inner sep=2em}}
\tikzset{line/.style={draw, thick, -latex'}}
\usepackage{nicefrac}
\usepackage{caption}
\usepackage{pdflscape}
\usepackage{graphicx}
\usepackage{caption,booktabs}
\usepackage{amsmath}
\usepackage{amstext} 

\usepackage[abs]{overpic}

\usepackage[]{hyperref}
\hypersetup{
    colorlinks,%
    citecolor=blue,%
    filecolor=black,%
    linkcolor=blue,%
    urlcolor=black
}
\usepackage{subfig}
\usepackage{cleveref}
\usepackage{tabularx}
\usepackage{multicol}

 \usepackage[bold]{hhtensor}
\usepackage{algorithm}
\usepackage{algpseudocode}

\newlength\imagewidth
\newlength\imagescale

\biboptions{sort&compress}
 
\usepackage{scrextend}

\modulolinenumbers[1]

\begin{document}
 \tdplotsetmaincoords{0}{0}
 
\begin{frontmatter} 

\title{Thermal super-jogs control high-temperature strength in Nb-Mo-Ta-W alloys}

\author[UCLA_MSE]{Sicong He}
\author[UCLA_MSE]{Xinran Zhou}
\author[TECH]{Dan Mordehai}
\author[UCLA_MSE,UCLA_MAE]{Jaime Marian}
\ead{jmarian@ucla.edu}
 
\address[UCLA_MSE]{Department of Materials Science and Engineering, University of California, Los Angeles, CA 90095, USA}
\address[TECH]{Department of Mechanical Engineering, Technion-Israel Institute of Technology, Haifa 3200003, Israel}
\address[UCLA_MAE]{Department of Mechanical and Aerospace Engineering, University of California, Los Angeles, CA 90095, USA}
 
\begin{abstract}
Refractory multi-element alloys (RMEA) with body-centered cubic (bcc) structure have been the object of much research over the last decade due to their high potential as candidate materials for high-temperature applications. Most of these alloys display a remarkable strength at temperatures above 1000$^\circ$C, which cannot be explained by the standard model of bcc plasticity dominated by thermally-activated screw dislocation motion. Recent research on Nb-Mo-Ta-W alloys points to a heightened role of edge dislocations during mechanical deformation, which is generally attributed to atomic-level chemical fluctuations in the material and their interactions with dislocation cores during slip. However, while this effect accounts for levels of strength that are much larger than what might be found in a pure metal, it is not sufficient to explain the high yield stress found at high temperature in Nb-Mo-Ta-W. In this work, we propose a new strengthening mechanism based on the existence of thermal super-jogs in edge dislocation lines that act as strong obstacles to dislocation motion, conferring an extra strength to the alloy that turns out to be in very good agreement with experimental measurements. The basis for the formation of these super-jogs is found in the unique properties of RMEA, which display vacancy formation energy distributions with tails that extend into negative values. This leads to spontaneous, i.e., athermal, vacancy formation at edge dislocation cores, which subsequently relax into atomic-sized super-jogs on the dislocation line. At the same time, these super-jogs can displace diffusively along the glide direction, relieving with their motion some of the extra stress, thus countering the hardening effect due to jog-pinning
We implement these mechanisms into a specially-designed hybrid kinetic Monte Carlo/Discrete Dislocation Dynamics approach (kMC/DD) parameterized with vacancy formation and migration energy distributions obtained with machine-learning potentials designed specifically for the Nb-Mo-Ta-W system. The kMC module sets the timescale dictated by thermally-activated events, while the DD module relaxes the dislocation line configuration in between events in accordance with the applied stress. We find that the balance between super-jog pinning and super-jog diffusion confers an extra strength to edge dislocations at intermediate-to-high temperatures that is in remarkable agreement with experimental measurements in equiatomic Nb-Mo-Ta-W and several other RMEA. 
We derive an analytical model based on the computational results that captures this improved understanding of plastic processes in these alloys and  and explains the experimental data.
\end{abstract}


\end{frontmatter}
 
\section{Introduction}

Since their inception in the 1990s \cite{CANTOR2004213,doi:10.1002/adem.200300567}, high entropy alloys (HEA) have attracted a great deal of attention due to a unique combination of properties seldom found in other material types \cite{zhang2013,yeh2007,doi:10.1080/09506608.2016.1180020,RN14293,MIRACLE2017448,doi:10.1080/21663831.2014.912690,senkov_miracle_chaput_couzinie_2018,met8020108,Gludovatz1153}. This makes them potentially very attractive as candidate materials for a number of technological applications in harsh environments such as elevated temperatures, irradiation, or corrosion \cite{xia2015irradiation,xia2016phase,kumar2016microstructural,egami2014irradiation}. 
The basic idea behind creating alloys of this type is to combine a number of elements (typically four or more) in similar proportions to achieve solid solution phase stability through the large configurational entropy of the system. Due to the large chemical and configurational space available to create these materials, several hundred different HEA combinations now exist, each with their own distinct compositions, structure, and properties \cite{Singlephasehighentropyalloysanoverview,10.1115/ETAM2018-6732,senkov2015accelerated}. The large volume of research on the topic over the last decade has resulted in a fast-evolving field full of new findings, unexplained results, and unresolved controversies. The reader is referred to the numerous reviews and monographs published over the last several years for more details \cite{zhang2013,yeh2007,doi:10.1080/09506608.2016.1180020,RN14293,MIRACLE2017448,doi:10.1080/21663831.2014.912690,senkov_miracle_chaput_couzinie_2018,met8020108,e16010494,f0c336970db7455aadf17c8930f61be6}. Note that, while the term `multicomponent' or `multielement' alloys is sometimes preferred in the literature over `high-entropy' alloys (particularly when the number of elemental constituents is less than five), here we use all interchangeably.

Among the different materials proposed, refractory multi-element alloys (RMEA) are a special class of alloys composed of typically four or more refractory metal elements (Nb, Mo, Ta, V, W, Cr, Hf, Zr). These systems generally crystallize into a single body-centered cubic (bcc) phase, found to be stable up to very high temperatures \cite{senkov_miracle_chaput_couzinie_2018,SENKOV2011698,ZOU201485,e18050189,e18080403,YAO20171139,DOBBELSTEIN2016624,KUBE2019677}. RMEA display sluggish self-diffusion rates  \cite{chang2014,beke2015,ZHAO2017391,ROY2021107106} and, similar to their pure bcc metal counterparts, suffer from a lack of ductility at low temperatures \cite{SENKOV2011698,RN142917}. However, they retain a high strength and ductility at high temperature, making them attractive candidates for structural applications in the power, aerospace, or nuclear sector \cite{xia2015irradiation,xia2016phase,kumar2016microstructural,yeh_lin_2018,e16010494}. 

While the deformation mechanisms of bcc metals and their alloys are relatively well understood, the mechanical response of chemically complex alloys at high temperature is still under vigorous investigation. Perhaps one of the most crucial aspects separating standard bcc behavior with that of RMEA is the role played by screw and edge dislocations during plastic deformation. Although the experimental evidence is still inconclusive \citep{couzinie2015room,couzinie2019body,bu2021local}, recent research points to the increased importance of edge dislocation slip relative their role in pure metal bcc systems and dilute alloys \cite{chen2020unusual,lee2021strength,lee2020temperature,yin2021atomistic}. Because chemical fluctuations in HEA take place at the atomistic scale, molecular dynamics (MD) has become the preferred tool to simulate dislocation processes in these systems.
Indeed, MD simulations have been used to characterize the intrinsic roughness of dislocation lines \cite{rao2017atomistic,chen2020unusual,ma2020unusual}, the effect of short-range order (SRO) \cite{chen2020unusual,yin2021atomistic,schon2021short} on alloy properties, and dislocation mobilities \cite{yin2021atomistic,shen2021mobility}. For example, MD simulations have convincingly shown that the lattice resistance aided by SRO (intrinsic strengthening) in bcc RMEA can reach up to 1 GPa at low temperatures. However, their contribution is much diminished as the temperature increases, and it alone cannot account for the observed high-temperature dependence of alloy strength.

In contrast, an often overlooked feature of multi-element systems is the statistical nature of defect properties due to their compositional heterogeneity \cite{wang2017thermodynamics,ding2019tuning,PhysRevMaterials.5.103604,PhysRevB.104.104101,roy2022vacancy}. Properties such as point defect formation energies, planar defect energies, or dislocation core energies are defined by distributions whose variance generally correlates with the number of elements in the alloy \cite{PhysRevMaterials.5.103604,PhysRevB.104.104101}. In particular, the thermal concentration of point defects in pure refractory metals is typically negligible on account of their relatively high formation energies. However, in HEA the low energy tails of defect energy distributions can lead to non-insignificant defect concentrations, even in conditions where these would be several orders of magnitude lower if one considered only the individual alloy elements. Moreover, defect formation energies can be substantially reduced at material inhomogeneities such as dislocation cores, material interfaces, precipitates, etc.~\cite{kabir2010predicting,HOSSAIN2014107}. For vacancies, which display formation energies 2$\sim$3 times lower than self-interstitial atoms in metals, a side effect of above two effects combined (broad energy distributions together with heterogeneous defect nucleation), is the possibility for non-thermal formation, i.e., the spontaneous existence of vacancies at any temperature at pre-existing microstructural defects.

Of particular interest is the formation of vacancies at or near edge dislocation cores, which is thermodynamically more favorable within the compressive region of the stress field \cite{HOSSAIN2014107}. Topologically, vacancies on dislocation lines can be regarded as elementary (super)jogs that can only move by nonconservative means (e.g., \emph{jog dragging}) \cite{zehetbauer1994effects,louchet2000ordinary,rao2021theory}. As such, the existence of super-jogs on a dislocation line can lead to extra strengthening, as has been seen in a number of systems at high temperature \cite{feng1999cross}. Moreover, owing to the low energy tails of the vacancy formation energy distributions, it is reasonable to expect a nonzero concentration of super-jogs on edge dislocation lines in even at low temperatures. The main objective of this work is to quantify this specific effect in a model RMEA and establish its viability as a high-temperature strengthening mechanism. More generally, we aim to study the behavior of edge dislocations, and their impact on alloy strength, when single-valued point defect energies are replaced by energy distributions.

Due to MD's intrinsic limitations when it comes to simulating thermally-activated mechanisms and their role in dislocation motion, here, following our line of work in previous publications  \citep{zhou2021cross,zhou2021temperature}, we conduct stress-driven simulations of edge dislocation motion as a function of temperature using a combined approach consisting of discrete dislocation dynamics (DD) and kinetic Monte Carlo (kMC). Our DD/kMC model captures the thermal formation and evolution of vacancies/super-jogs on edge dislocations and quantifies their effect on dislocation motion as a first attempt to embed some of the complexites of the highly-random alloy into a mesoscale framework with the aim of capturing time and length scales more suitable for experiments, as compared to MD. We focus on equiatomic Nb-Mo-Ta-W alloys, for which robust interatomic potentials exist \cite{zhou2019machine,li2020complex,rickman2020machine}, as well as a wealth of experimental data \cite{doi:10.1080/21663831.2016.1198837,FENG2017264,e18080403,xin2018ultrahard,zhang2021high,marta-hea-2022}, and carry out calculations of vacancy energetics to parameterize the model without fitting to any experimental measurements.

The paper is organized as follows. First, we provide the theoretical framework for DD/kMC model. This is followed by a description of the numerical implementation and the atomistic calculations performed to parameterize the model to the alloy of choice. Next we present results of the strength of the alloy as a function of temperature and strain rate. We finalize the paper with a discussion of the results and the conclusions.

\section{Theory and models}

\subsection{Monte Carlo model of non-conservative edge dislocation motion}\label{monte}

The key element of the modeling approach is a stochastic model based on the residence-time algorithm to simulate dislocation evolution under stress. The time scale is governed by a set of thermally-activated events that punctuate periods of standard dislocation glide. As such, dislocation evolution proceeds by way of a series of super-jog processes intercalated by stress-assisted glide. Here we consider two distinct thermal processes, which are schematically illustrated in Figure \ref{translation}:
\begin{figure}[ht!]
   \centering
 \fbox{\subfloat[\label{ed1}]{\includegraphics[width=0.65\textwidth]{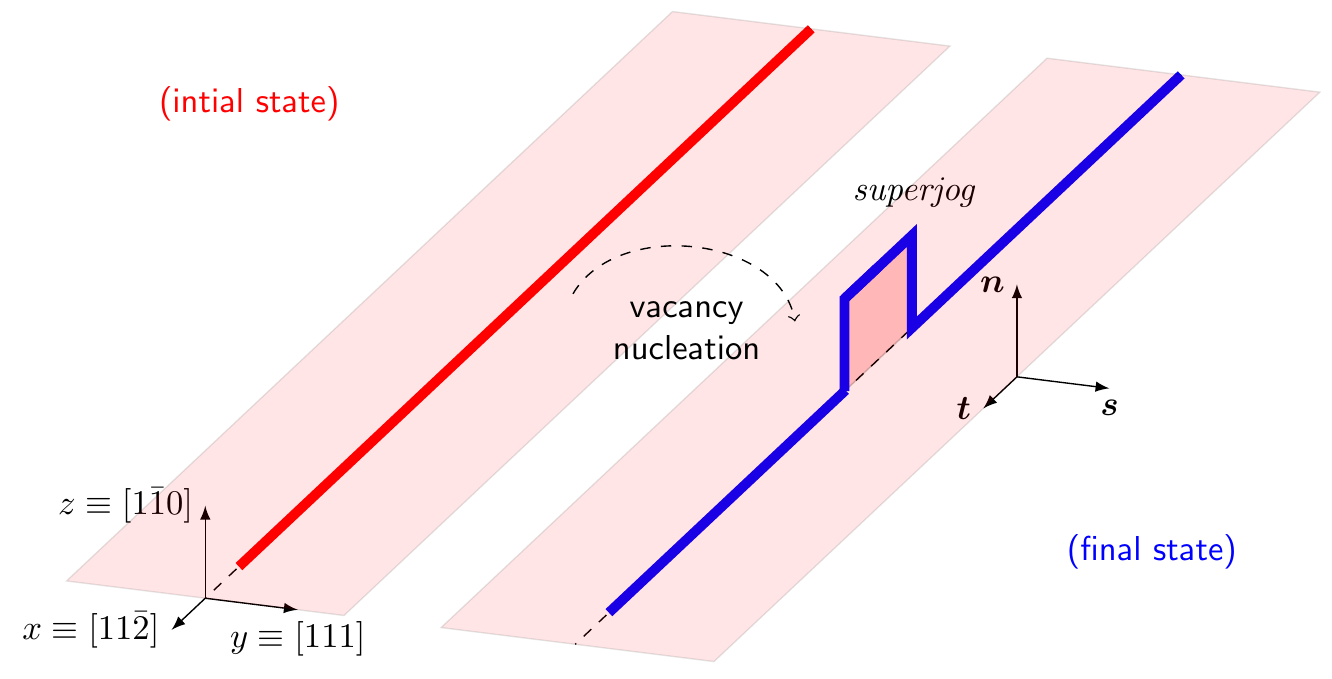}}}\\
  \fbox{\subfloat[\label{ed2}]{\includegraphics[width=0.65\textwidth]{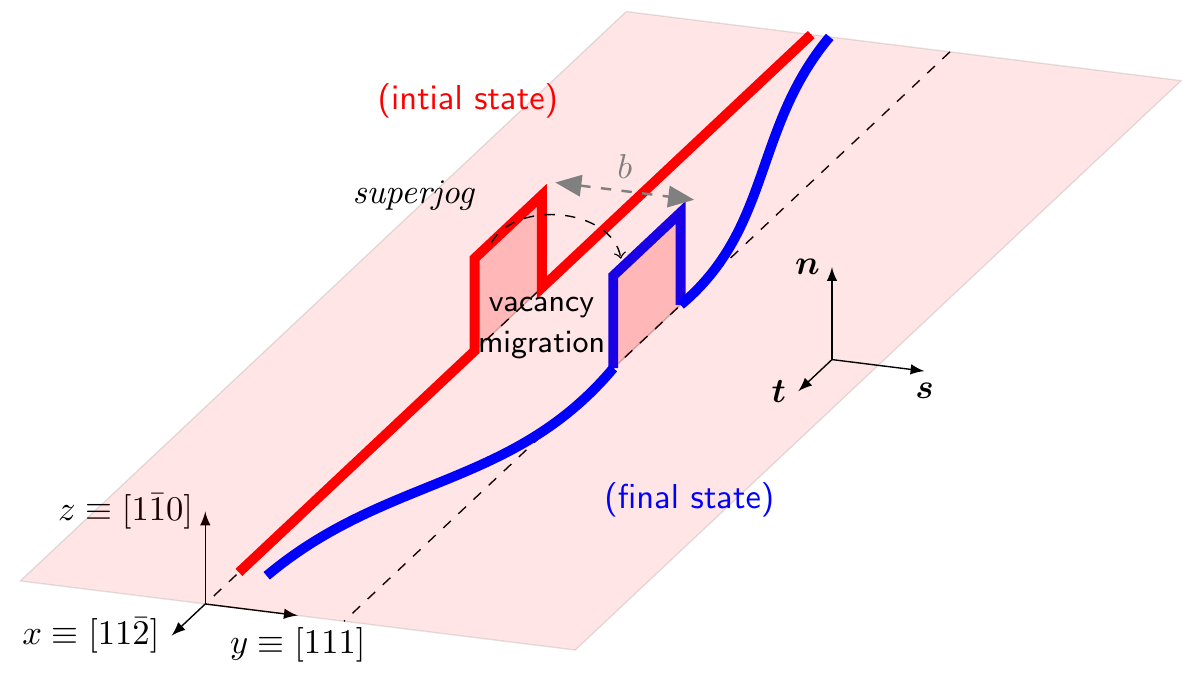}}}
   \caption{Modes of super-jog evolution. (a) super-jog generation on a straight dislocation line. This mode of motion requires vacancy nucleation. (b) Direct translation along the slip direction $\vec{s}$. This mode of motion requires a vacancy jump.}
   \label{translation}
\end{figure}
\begin{enumerate}
\item At any point during a simulation, a super-jog may appear on the dislocation line by thermal nucleation of a vacancy (process shown in Fig.\ \ref{ed1}). Such nucleation is subjected to equilibrium constraints so that the linear concentration of super-jogs remains within thermal limits. For this, the dislocation line is subdivided into irreducible segments of length $w=\frac{a_0\sqrt{6}}{3}$ (equal to the periodicity along the line direction $\nicefrac{1}{6}\langle112\rangle$), where $a_0$ is the lattice parameter. The insertion rate of  super-jogs per unit length is:
	\begin{equation}
	q^{\rm jog}_n(T;t)=\nu_{0}\left\lfloor\frac{\ell}{w}-n(t)\right\rfloor\left(1-\frac{n(t)}{n_{0}(T)}\right)
	\label{q3}
	\end{equation}
where $T$ is the absolute temperature, $\nu_{0}$ is a temperature-independent vacancy nucleation attempt frequency, $n(t)$ is the number of super-jogs at time $t$ (the current number of super-jogs), and $n_{0}(T)$ is the equilibrium super-jog concentration (to be defined in Section \ref{param}). 
Here, the ratio $\ell/w$ represents the number of potential nucleation sites on a dislocation with total length $\ell$ for a super-jog to form, while $\left\lfloor\ell/w-n(t)\right\rfloor>0$ represents the number of available nucleation sites\footnote{The function $\lfloor x \rfloor$ is the mathematical \texttt{floor} function.}. The `height' of a jog segment is the unit lattice parameter along the $y$-direction: $h=\left\|\nicefrac{1}{2}[110]\right\|=a_0\nicefrac{\sqrt{2}}{2}$.
Note that, while thermodynamically possible, super-jog reabsorption is highly unlikely because it requires either the arrival of self-interstitial atoms or emission of vacancies, both of which are discarded in this model due to their high penalizing energies.

\item Existing super-jogs can translate along the glide $x$-direction via diffusive jumps to a location separated one Burgers vector's distance from the original position (Fig.\ \ref{ed2}). This process is akin to the jump of a vacancy located on the edge dislocation core along the glide $[111]$ direction, which corresponds to a nearest-neighbor jump in a bcc lattice. The rate of this process can be written as:  
	\begin{equation}
	q^{\rm jog}_m(T)=\nu_{0}'\exp\left(-\frac{\Delta H_m^\perp+\Delta E_{\rm el}^{\rm jog}}{kT}\right)
		\label{q1}
	\end{equation}
where $\nu_{0}'$ is the jump attempt frequency and $\Delta H_m^\perp$ is the vacancy migration enthalpy, which incorporates contributions from the stress state at the super-jog location, and $\Delta E_{\rm el}^{\rm jog}$ is the extra elastic energy incurred by the bending of dislocation segments adjacent to the super-jog (as shown in Fig.\ \ref{ed2}). Both of these quantities will be defined and calculated in Sec.\ \ref{param}. The above rate must be defined for both forward and backward jumps, with the direction of the jump assigned by defining the resolved shear stress as positive when it creates a force aligned with the glide direction. 

\end{enumerate}
A summary of the properties of each type of event is given in Table \ref{tab1}.
\begin{table}[h!]
  \begin{center}
    \caption{Summary of super-jog transitions.}
    \label{tab1}
    \begin{tabular}{|c|c|c|c|} 
    \hline
          \textbf{Segment} & \textbf{Rate} & \textbf{Definition} & \textbf{Distance} \\
          \hline
     super-jog (nucleation) &	$q^{\rm jog}_n$ & eq.\ \eqref{q3} & (super-jog dimensions: $h$$\times$$w$) \\
     super-jog (forward/backward) &	$q^{\rm jog}_m$ & eq.\ \eqref{q1} & $b$ \\
     \hline
    \end{tabular}
  \end{center}
\end{table}

Using kMC, the set of event rates is sampled at each iteration and the appropriate process is selected and executed. Time is evolved as a Poisson variate:
\begin{equation}
\delta t = -\frac{\log\eta}{R_t}
\end{equation}
where $\eta$ is a uniform random number, $R_t = \sum_{\alpha=n,m} \left( \sum_j {q_{\alpha}^{\rm jog}}\right)$, and the subindices $\alpha$ and $i$ apply to the type of transition (`$n$' nucleation, `$m$' migration) and to the number of existing super-jogs, respectively. In between thermal events, i.e., during a time $\delta t$, the dislocation evolves elastically in response to the existing stresses using a suitable DD model (explained in Sec.\ \ref{dd}).
A flow diagram of the numerical method is given in \ref{app2}, Figure \ref{diagram}.
				
\subsection{Dislocation dynamics model}\label{dd}

The DD model must be sensitive to the scale of discrete steps on the dislocation line to capture the formation of vacancy induced jogs. As such, the main difference between the dislocation dynamics approach used here and standard DD codes \cite{bulatov2004scalable} is that the positions of the nodes delimiting the jogs are constrained to have specific lengths. Beyond that, dislocation segments interact in an isotropic elastic manner with one another using the elastic constants of the alloy obtained with the potentials described in Sec.\ \ref{param}.

\begin{figure}[ht!]
   \centering
\includegraphics[width=0.65\textwidth]{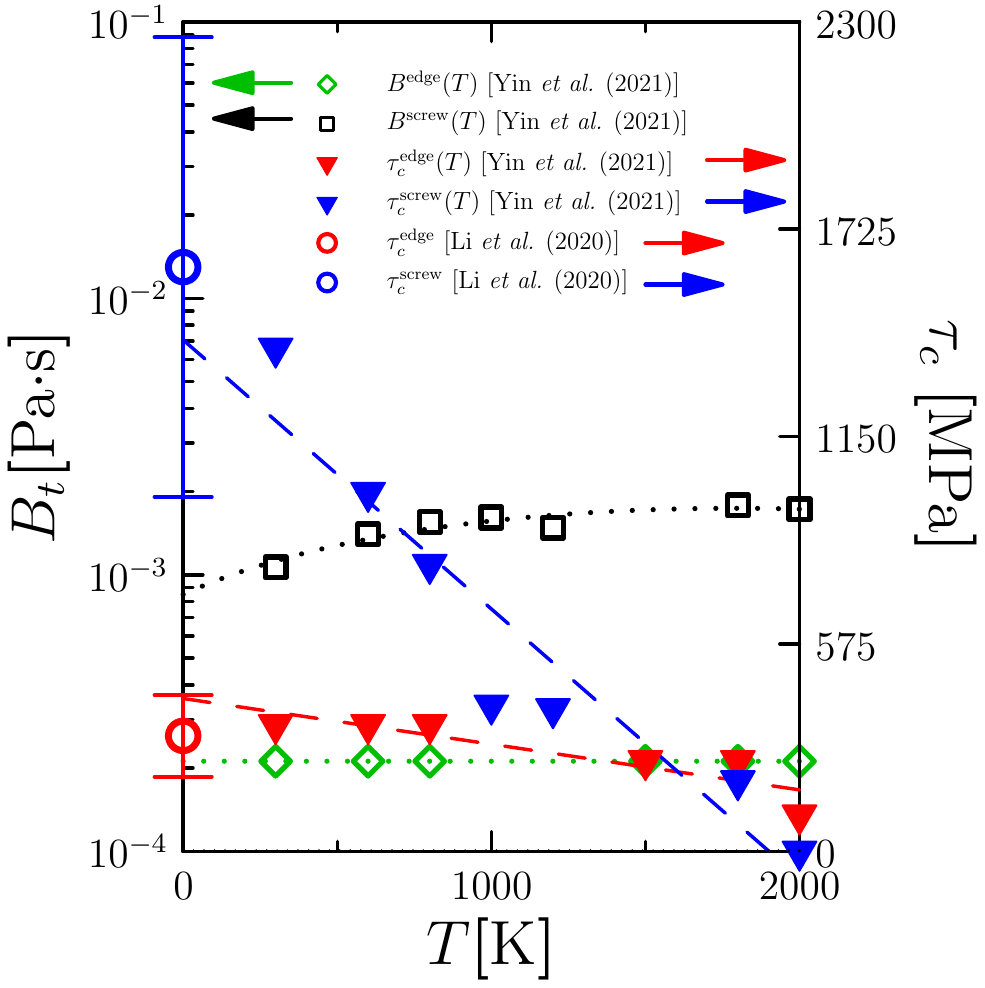}
   \caption{Dislocation critical stresses and friction coefficients from ref.\ \cite{yin2021atomistic}. The two data points for screw and edge at 0 K were obtained using a SNAP potential \cite{li2020complex}.}
   \label{crss}
\end{figure}

As is customary in most DD formulations, the segment glide velocity $v^{\rm gl}_i$ is linearly dependent on stress as:  
\begin{equation}
v^{\rm gl}=\frac{b\Delta \tau}{B_t(T,\theta)}\label{qglide}
\end{equation}
where $b=\|\vec{b}\|$ is the Burgers vector's modulus,  $\theta$ represents the dislocation character ($\cos\theta=\vec{s}\cdot\vec{t}$, where $\vec{s}=b^{-1}\vec{b}$ is the slip direction and $\vec{t}$ the local line tangent), $B_t$ is a temperature-dependent drag coefficient, and $\Delta \tau$ is the excess glide stress, obtained as the difference between the resolved shear stress $\tau_{\rm RSS}$ calculated at the segment or nodal position and a temperature and dislocation character dependent critical stress, $\tau_c$ which is a material constant. 
\begin{equation}
\Delta \tau=\tau_{\rm RSS}-\tau_c(T,\theta)
\label{Deltatau}
\end{equation}
with:
\begin{equation}
\tau_c(T,\theta)=\tau^{\rm screw}_c(T)\cos^2\theta+\tau^{\rm edge}_c(T)\sin^2\theta
\end{equation}
Likewise, for the drag coefficient $B_t$, the same dependences on $T$ and $\theta$ are used:
\begin{equation}
B_t(T,\theta)=B^{\rm screw}(T)\cos^2\theta+B^{\rm edge}(T)\sin^2\theta
\end{equation}
The parameters $\tau^{\rm screw}_c(T)$, $\tau^{\rm edge}_c(T)$, $B^{\rm screw}_c(T)$, and $B^{\rm edge}_c(T)$ are temperature dependent critical stresses and glide friction coefficients for screw and edge coefficients, respectively.
Here we employ coefficients derived from recent results by Yin \etal~\cite{yin2021atomistic} for the Nb-Mo-Ta-W equiatomic system, shown in Figure \ref{crss}. The numerical expressions for the data in the figure used in the present DD simulations are provided in Table \ref{tab:param}.
\begin{table}[h!]
\caption{Numerical expressions use for the temperature dependence of the coefficients plotted in Fig.\ \ref{crss}. $T$ is the absolute temperature in K.\label{tab:param}}
\begin{center}
\begin{tabular}{|c|c|c|}
\hline
Parameter & Expression & Units \\
\hline
&&\\[-1em]
$B^{\rm screw}(T)$ & $3.44\times10^{-4}\ln(T) -8.37\times10^{-4}$ & Pa$\cdot$s \\
$B^{\rm edge}(T)$ & $2.12\times10^{-4}$ & Pa$\cdot$s \\
\hline
$\tau^{\rm screw}_c(T)$ & $1418.0 - 0.7T$ & MPa \\ 
$\tau^{\rm edge}_c(T)$ & $423.6 - 0.1T$ & MPa \\
\hline
\end{tabular}
\end{center}
\label{default}
\end{table}%
Here we use a shear modulus of $\mu=94$ GPa and a Poisson ratio of $\nu=0.33$ \cite{li2020complex}.

The geometry used in the DD simulations follows a Cartesian coordinate system with axes $x\equiv[11\bar{2}]$, $y\equiv[111]$, and $z\equiv[1\bar{1}0]$ representing, respectively, the dislocation line direction, the plane normal, and the glide direction (along the Burgers vector $\vec{b}$). With such an orientation, an external stress tensor with $\tau_{yz}$ as the only nonzero component is applied and added to the total segment-segment elastic stresses at each point. While a shear stress $\tau_{yz}$ produces a net force on segments aligned with the $z$ direction, i.e., jog segments, we only allow slip on close-packed $\{110\}$ these segments and thus these segments cannot glide conservatively during the simulations. They can, however, move non-conservatively, which as will be shown is dealt with by the kMC module.

\subsection{Model parameterization}\label{param}

\subsubsection{Atomistic calculations of vacancy formation and migration energies}

The two main processes described earlier are characterized by event rates that involve vacancy transitions at or next to edge dislocation cores. The first one, thermal nucleation of super-jogs, is governed by the vacancy formation energy at edge dislocation cores, $\Delta E_f^{\rm V@\perp}$. 
The second one, super-jog translation along the glide direction, is represented by the migration energy along the glide direction of vacancies lying on edge dislocation cores, $\Delta E_m^{0\rightarrow1}$.  For simplicity going forward, we use shorthand notation to refer to $\Delta E_f^{\rm V@\perp}$ and $\Delta E_m^{0\rightarrow1}$, as $E^{\perp}_f$ and $E^{\perp}_m$, respectively.

To calculate the distribution of $E^{\perp}_f$ and $E^{\perp}_m$, we first construct an edge dislocation dipole in an atomistic supercell using the same procedure described in detail by Hossain and Marian \cite{hossain2014stress}. 
In this work we use a spectral neighbor analysis potential (SNAP) for the Nb-Mo-Ta-W system, which has been trained against a comprehensive materials database \cite{snap,li2020complex}. A value of $a_0=3.24$~\AA~is obtained for the lattice parameter of the alloy using this potential.

\begin{figure}[h!]
   \centering
   \subfloat[\label{edcna}]{
   \includegraphics[width=0.5\textwidth]{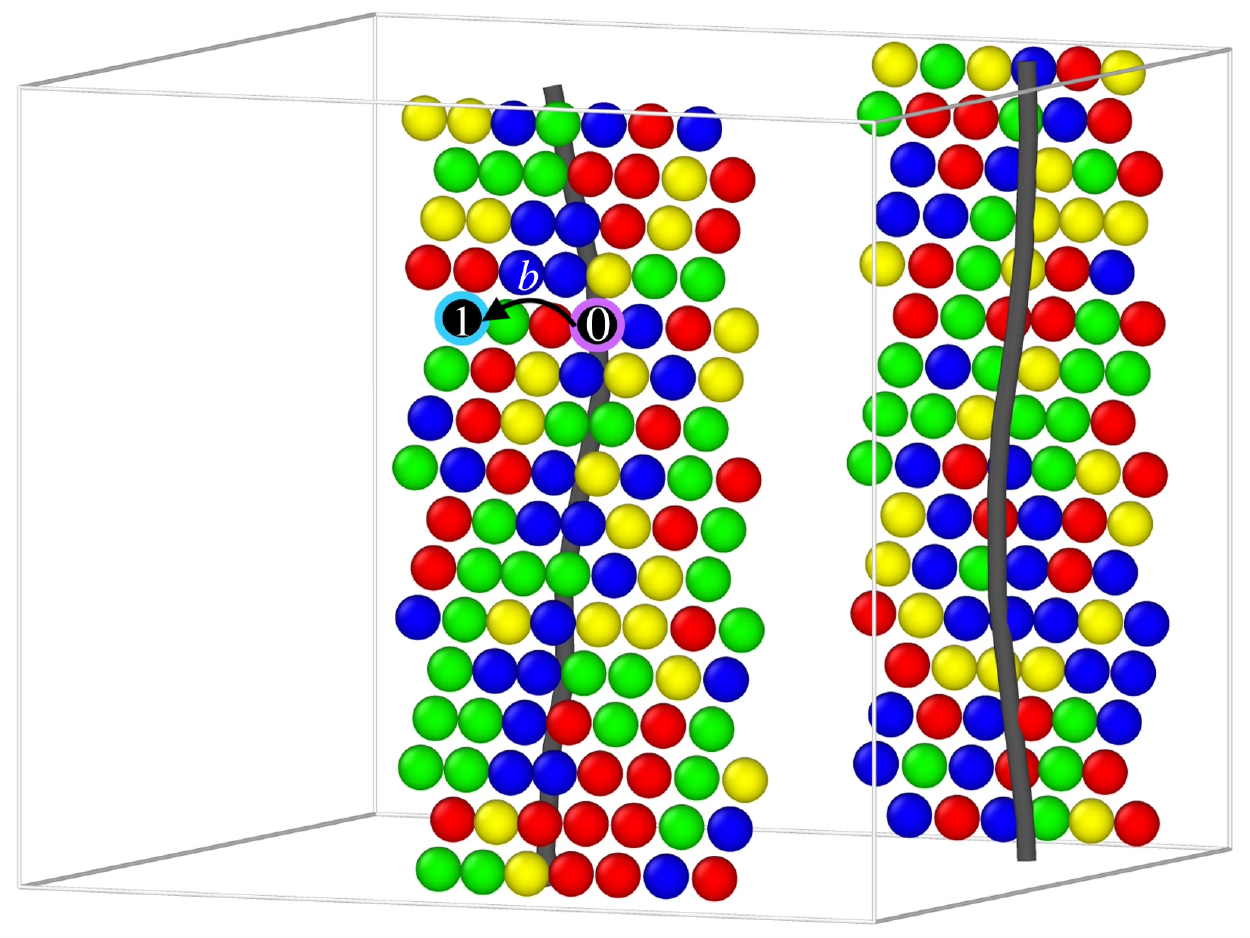}
   }
   \subfloat[\label{edcolor}]{
   \includegraphics[width=0.5\textwidth]{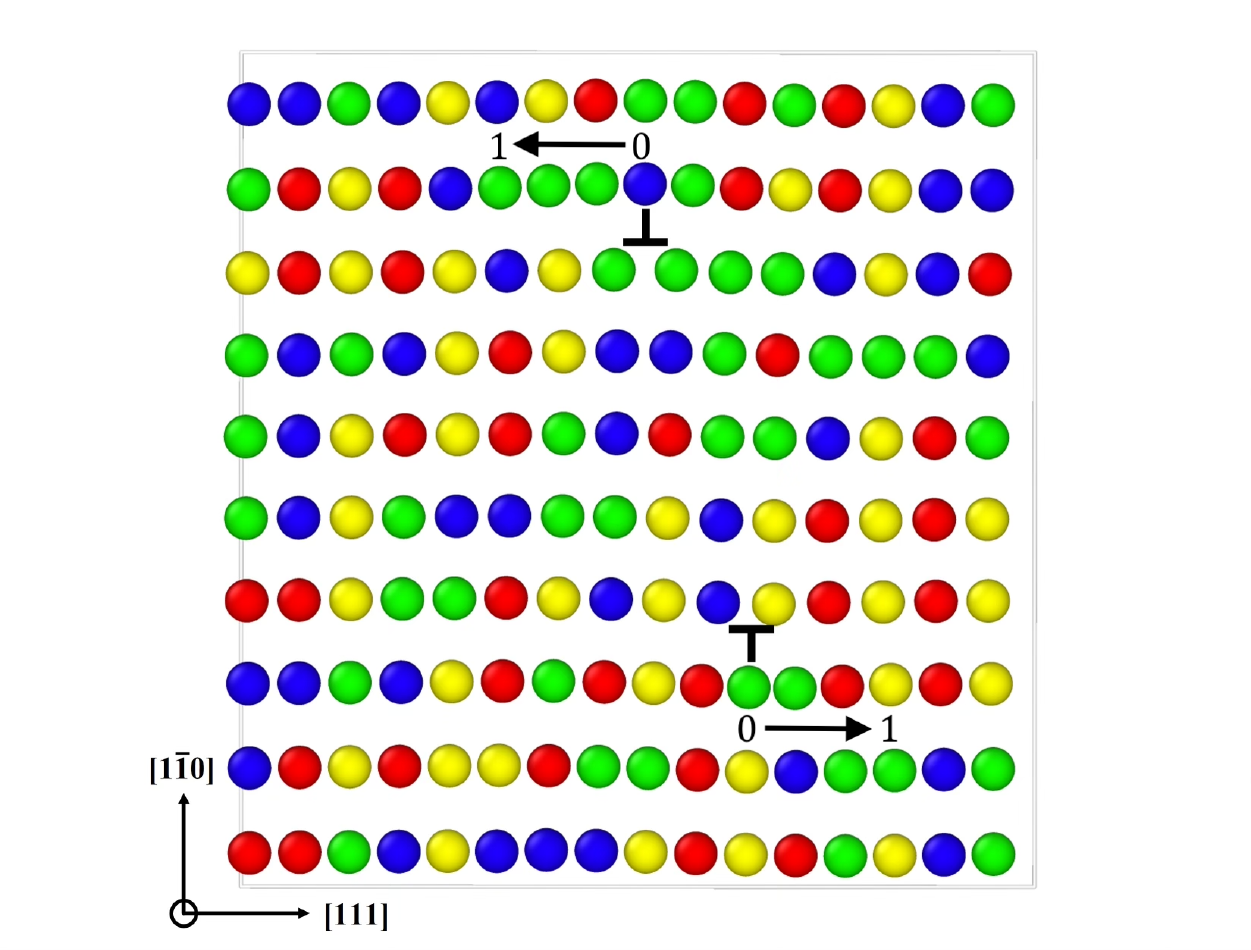}
   }
   \caption{Relaxed edge dislocation dipole utilized for vacancy formation energy calculations. (a) Atomistic representation according to common neighbor analysis and equivalent line representation, both extractyed using \texttt{Ovito} \cite{ovito}. (b) Side view showing the edge pole dislocation locations (`$\perp$' symbols) and atoms from three consecutive $\left[11\bar{2}\right]$ planes.  Atoms are colored by chemical species (green for Nb, red for Mo, blue for Ta, and yellow for W). The atomic positions `0' and `1' (of significance in the text) are indiocated in both images. The distance between positions 0 and 1 is equal to the first nearest neighbor distance (i.e., one Burgers vector) with both positions located on the same $\left[11\bar{2}\right]$ plane.}
   \label{edge1}
\end{figure}

Figure \ref{edcna} shows the relaxed atomic structure of the dipole visualized according to common neighbor analysis and the dislocation extraction algorithm using \texttt{Ovito} \cite{ovito}, while Fig.\ \ref{edcolor} shows a side view with the two edge dislocation locations (`$\perp$' symbols) and two atomic positions, labeled `0' and `1', at and near the dislocation cores. 
$E^{\perp}_f$ is calculated at different `0' locations along the dislocation line. Note that while these locations are crystallographically equivalent, they are `chemically' different, which is the source of the variability in $E^{\perp}_f$. 

\begin{figure}[ht!]
   \centering
  \subfloat[\label{edpos}]{\includegraphics[width=0.40\textwidth]{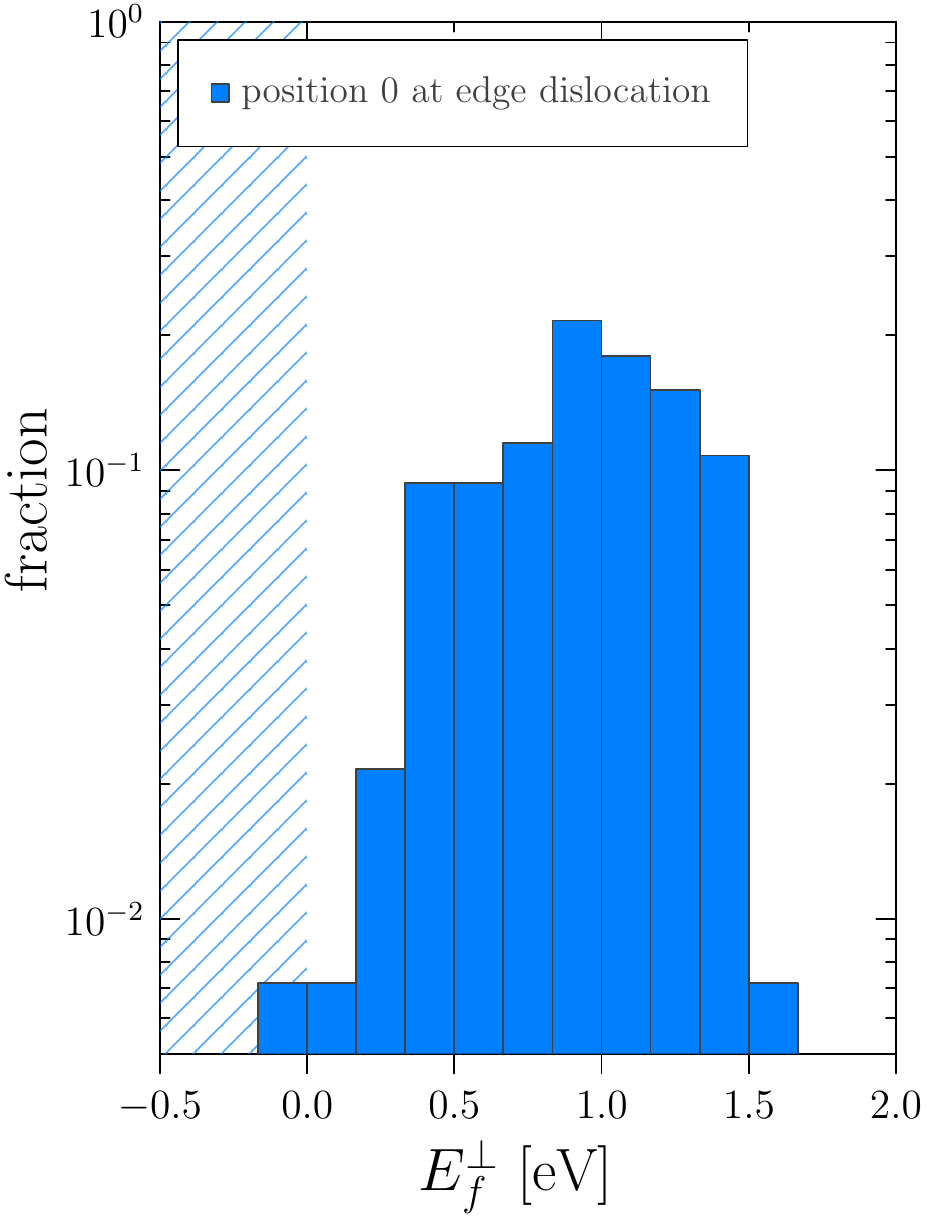}} 
  \subfloat[\label{edform}]{\includegraphics[width=0.268\textwidth,trim=0pt -3pt 0pt 0pt, clip]{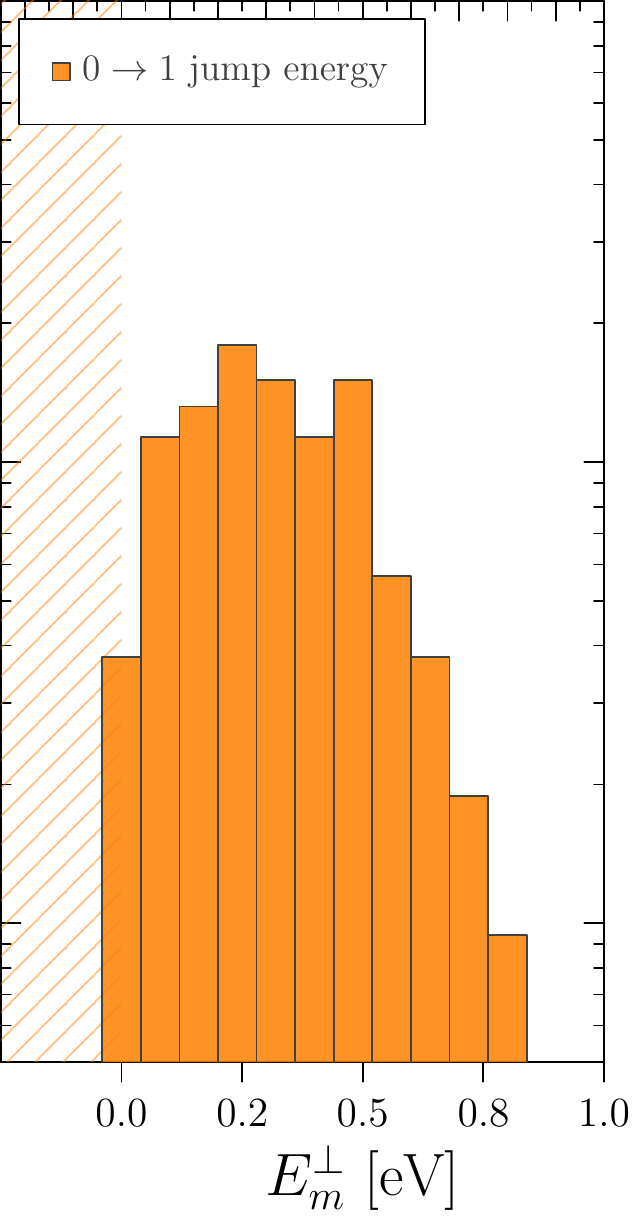}}
   \caption{(a) Vacancy formation energy distribution in the Nb-Mo-Ta-W alloy at position `0' near the edge dislocation core in Fig.\ \ref{edcolor}. (b) Migration energy distribution for the 1$\rightarrow$0 jump at the edge dislocation core positions shown in Fig.\ \ref{edpos}. Negative energies (shaded area in the graphs) signal the spontaneous formation of vacancies and transitions, respectively.}
   \label{edge0}
\end{figure}

For its part, $E^{\perp}_m$ is calculated as the migration energy between points `0' and `1', which are separated by an amount equal to the first nearest neighbor distance (equivalent to the Burgers vector's modulus) with both positions located on the same $\left[11\bar{2}\right]$ plane. The corresponding normalized distributions, $p(E^{\perp}_f)$ and $p(E^{\perp}_m)$, are provided in Figure \ref{edge0}.   

As a point of comparison, the statistical means of $p(E^{\perp}_f)$ and $p(E^{\perp}_m)$, denoted by $\bar{E}^{\perp}_f$ and $\bar{E}^{\perp}_m$ are tabulated here for the random Nb-Mo-Ta-W system along  with the corresponding formation and migration energies in bulk Nb-Mo-Ta-W and in the gray material.
\begin{table}[h!]
\begin{center}
\caption{Mean energies for the vacancy formation and migration energy distributions at dislocations cores (`$\perp$', highlighted in red) and in the bulk (`V'), for both the random Nb-Mo-Ta-W system and the \emph{gray} alloy. The relevant energies are highlighted in red.\label{tab:ener}}
\begin{tabular}{|c|c|c|}
\hline
Energies [eV] & Nb-Mo-Ta-W & `gray' alloy \\
\hline
&&\\[-1em]
$\bar{E}^{\perp}_f$ & \textcolor{red}{0.93} & -- \\ 
&&\\[-1em]
$\bar{E}^{\rm V}_f$ & 2.40 & 2.62 \\
\hline
&&\\[-1em]
$\bar{E}^{\perp}_m$ &  \textcolor{red}{0.32} & -- \\
&&\\[-1em]
$\bar{E}^{\rm V}_m$ & 1.71 & 1.53 \\
\hline
\end{tabular}
\end{center}
\end{table}
As Table \ref{tab:ener} shows, formation energies are substantially lower than in the bulk (averages of 0.93 versus 2.40 eV). Critically, however, some of the calculated energies are zero or negative, introducing the possibility for spontaneous vacancy formation. In other words, thermal vacancies are expected to form with considerable ease at edge dislocation cores compared to bulk positions, but, most importantly, in certain locations vacancies will form athermally to lower the total local configurational energy around the dislocation. This essentially implies that super-jogs, which are the manifestation of monovacancies in terms of dislocation lines, may naturally exist along the edge dislocation in equilibrium conditions at any temperature.

\subsubsection{Activation enthalpies and attempt rates}\label{enthal}

Next we turn our attention to the parameters in eqs.\ \eqref{q3} and \eqref{q1}. The key quantity to define in eq.\ \eqref{q3} is the thermal concentration of super-jogs, $n_0(T)$. In general, $n_0(T)=n_0^\ast\exp\left(-E^{\perp}_f/kT\right)$, where $n_0^\ast$ is a pre-factor determined by the crystal geometry. In our case, $n_0^\ast$ is set by the inverse of the  periodicity along the the dislocation line direction, which is equal to $1/w$.

Similarly, the jump rate in eq.\ \eqref{q1} can be expressed as:
\begin{equation}
q^{\rm jog}_m(T)=2\nu_{0}'\exp\left(-\frac{\Delta E^\perp_m+\Delta E_{\rm el}^{\rm jog}}{kT}\right)\sinh\left(\frac{\Delta\tau~\left(wb^2\right)}{kT}\right)
\label{sinh}
\end{equation}
This expression captures forward and backward jumps\footnote{Forward jumps are taken as those for which the stress tensor projected along the Burgers vector's direction results in a positive resolved shear stress along the glide direction, i.e., when $\left(\matr{\sigma}:\vec{b}\right)\cdot\vec{s}>0$.} through the $\sinh$ term, whose argument is the work done by the excess stress (refer to eq.\ \eqref{Deltatau}) to move the super-jog a distance $b$. The area swept during the process is equal to the width of the super-jog times $b$, i.e., $wb$.  The product $wb^2$ comes out to approximately $1.2\Omega_a$ ($\Omega_a=a_0^3/2$).
$\Delta E_{\rm el}^{\rm jog}$ is the extra elastic energy due to the bending of the dislocation segments adjacent to the super-jog, which is obtained directly by the DD module of the code.

With respect to the attempt frequencies $\nu_0$ and $\nu_0'$, here we adopt a  value of $10^{12}$ Hz for both. 

\section{Results}

\subsection{Thermal super-jog concentration}

\begin{figure}[h]
   \centering
\includegraphics[width=0.7\textwidth]{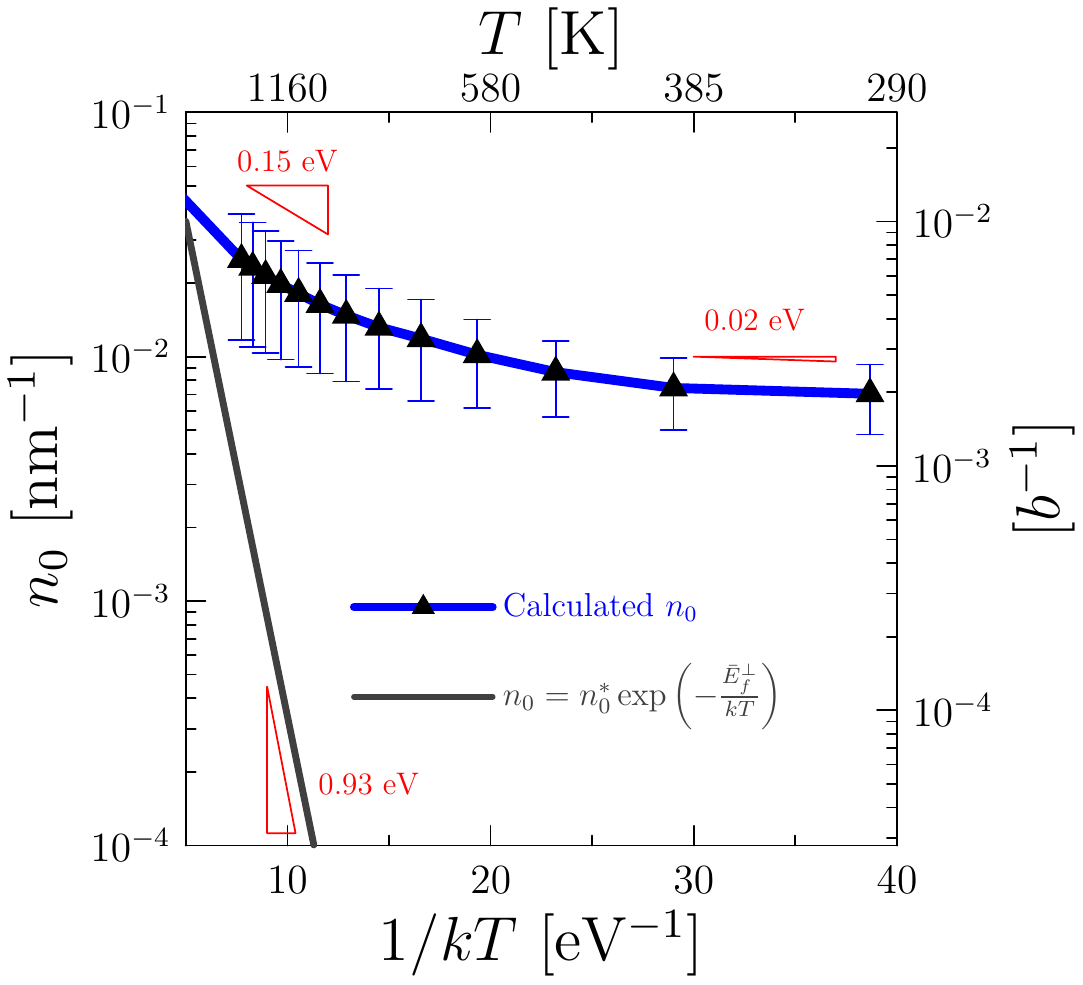}
   \caption{Arrhenius plot of the thermal concentration of super-jogs, $n_0$, as a function of temperature on edge dislocations. The figure includes both the lower-bound $n_0$, characterized by the mean energy of the formation energy distribution in Fig.\ \ref{edpos}, $\bar{E}^{\perp}_f=0.93$ eV, and also the simulated $n_0$, obtained by sampling from the distribution in Fig.\ \ref{edpos} according to eq.\ \eqref{q3} (obtained from 10 independent samplings at each temperature).}
   \label{n0}
\end{figure}

Figure \ref{n0} shows an Arrhenius plot of the thermal concentration of super-jogs, $n_0$, as a function of temperature in edge dislocations. The figure includes the lower-bound $n_0$, characterized by the mean energy of the formation energy distribution in Fig.\ \ref{edpos}, given in Table \ref{tab:ener} as $\bar{E}^{\perp}_f=0.93$ eV, and also the simulated $n_0$, obtained by sampling from the distribution in Fig.\ \ref{edpos} according to eq.\ \eqref{q3} (obtained from 10 independent samplings at each temperature). As the data show, the \emph{effective} formation energies are much lower than 0.93 eV, ranging from 0.02 eV below 400 K to 0.15 eV above 1000 K. Both distributions converge to the common prefactor of $n_0^\ast=1/w$ at very high temperatures. Overall the simulated thermal concentration of super-jogs is orders of magnitude larger than that given by the Arrhenius form of $n_0$. Moreover, it displays an almost athermal dependence with $T$, particularly at low temperatures. This is yet another manifestation of the unique properties of compositionally-complex alloys such as Nb-Mo-Ta-W, which display enhanced thermal defect concentrations due to asymmetric samplings of $p(E^{\perp}_f)$.

\subsection{Dislocation dynamics under stress}\label{ddd}

 Next we track dislocation motion under stress. We simulate a Frank-Read source with a fixed length $L$ pinned at its endpoints. $L$ is generally obtained as $L\approx\left(\rho_d\right)^{-1/2}$, so that its value is consistent with the dislocation density in the material, $\rho_d$. 
Values as low as $\rho_d=5\times10^{12}$ m$^{-2}$ have been reported in Nb-Mo-Ta-W \cite{zou2015ultrastrong}. However, this is three to four orders of magnitude smaller than dislocation densities measured in other bcc RMEA, such as $2\times10^{15}$ m$^{-2}$ in Ti-Nb-Hf-Ta \cite{thirathipviwat2020compositional}, $2.1\times10^{16}$ m$^{-2}$ in Hf-Nb-Ti-Zr \cite{gubicza2019evolution}, or $10^{15}$ m$^{-2}$ in V-Nb-Mo-Ta-W \cite{xin2018ultrahard}. Thus, we believe that the value of $\rho_d$ obtained by Zou {\it et al.}~for Nb-Mo-Ta-W \cite{zou2015ultrastrong} may be unrealistically low and, as such, we consider values of $1\sim2\times10^{15}$ m$^{-2}$ as being more representative of the dislocation density in the alloy. This gives rise to average dislocation source lengths on the order of 100$\sim$300$b$. Note that, in accordance with the results in Fig.\ \ref{n0}, these lengths are sufficient to contain at least one super-jog at temperatures above 580 K. The effect of $\rho_d$ on the model results is further discussed in Sec.~\ref{limit}. 

\subsection{Time evolution of the plastic strain}

\begin{figure}
    \centering
    \pgfmathsetlength{\imagewidth}{0.8\linewidth}%
    \pgfmathsetlength{\imagescale}{\imagewidth/524}%
    \begin{tikzpicture}[x=\imagescale,y=-\imagescale]
        \node[anchor=north west] at (-10,-7) {\includegraphics[width=0.8\imagewidth]{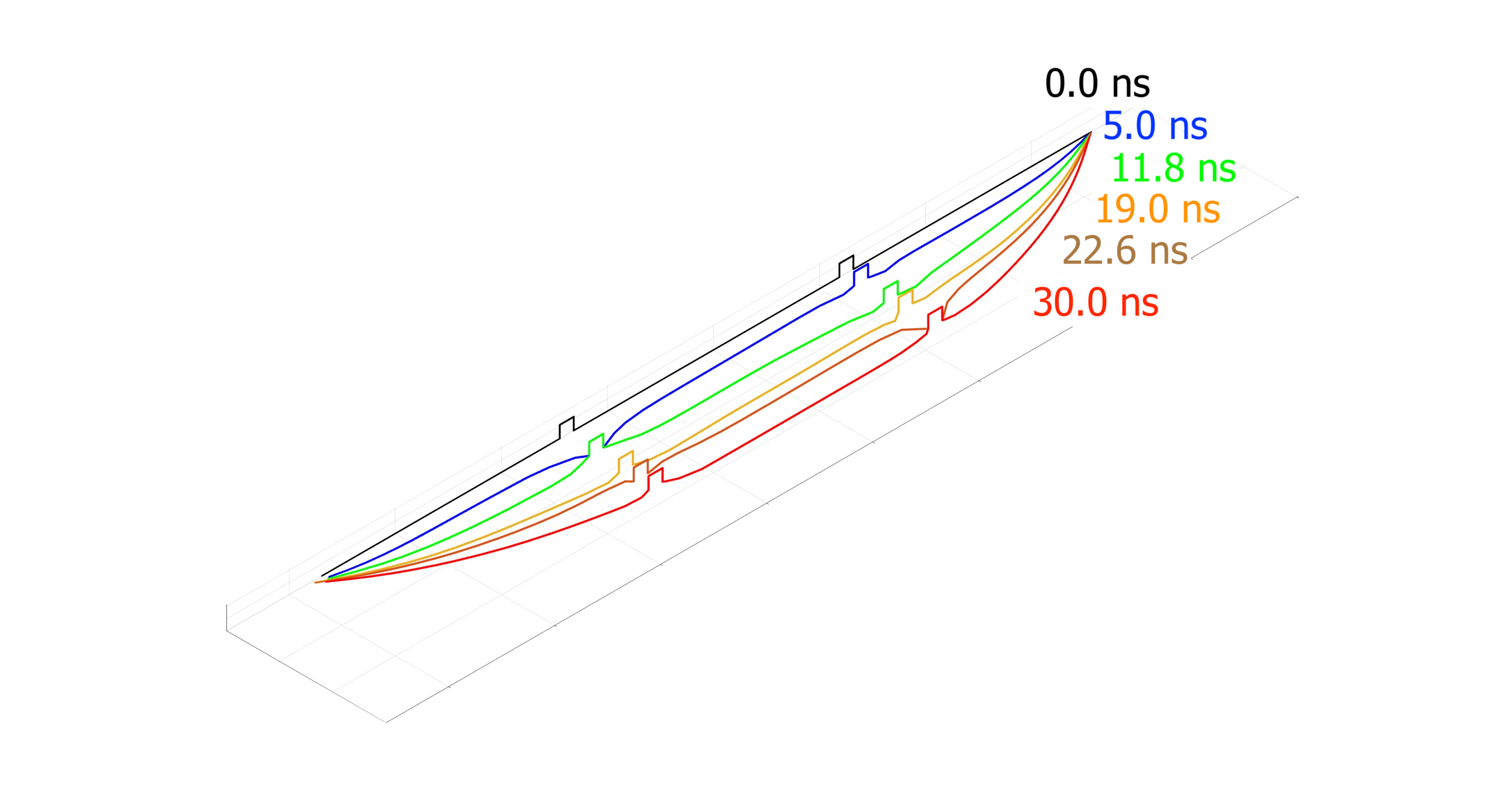}};
        \node[anchor=north west] at (0,0) {\includegraphics[width=0.85\imagewidth]{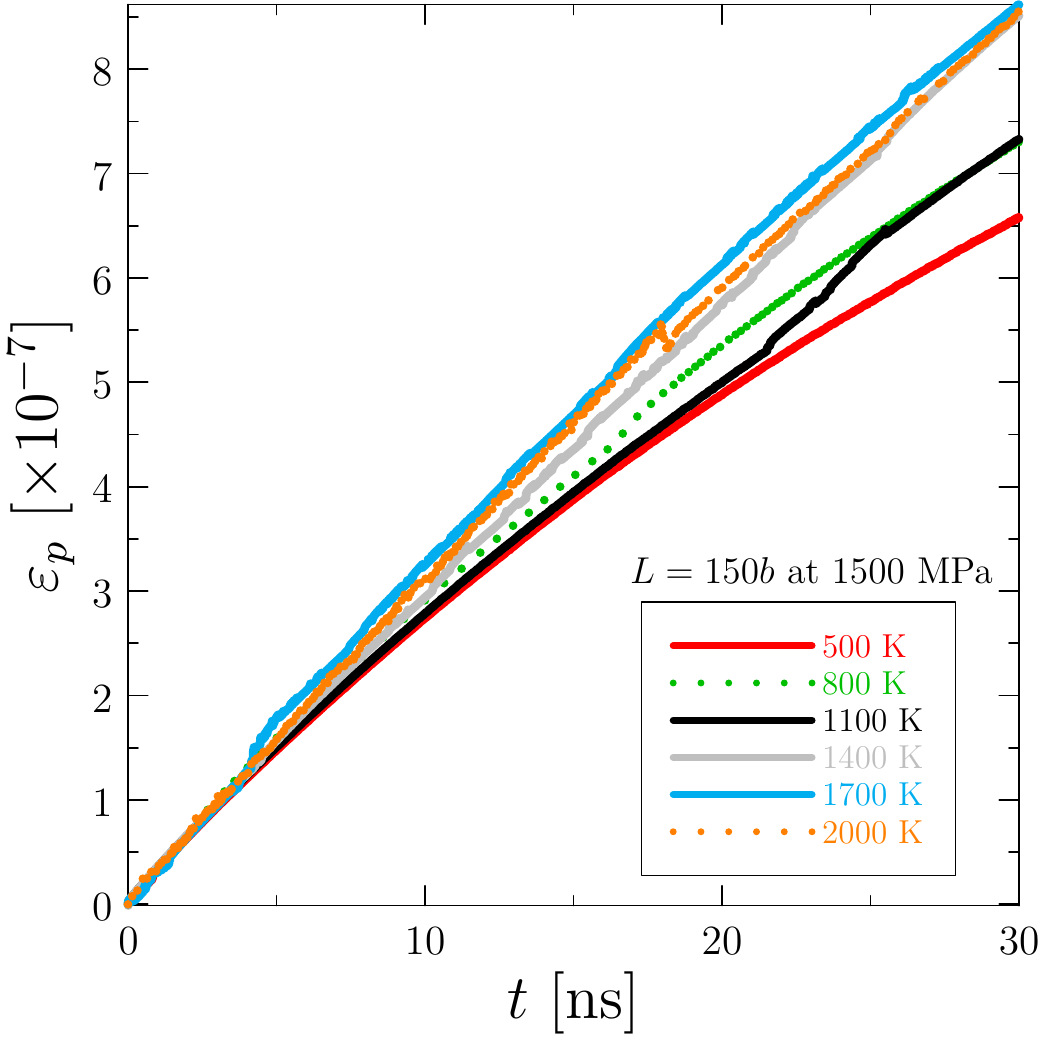}};
    \end{tikzpicture}
    \caption{Plastic strain rate as a function of time at several temperatures under an applied stress of 1500 MPa for a dislocation source with a total length of $150b$ containing two super-jogs evenly spread at $+L/3$ and $+2L/3$. The inset shows an overlapping sequence of time frames of the dislocation configuration at 1700 K covering 30 ns of DD/kMC simulation.}
\label{straintime}
\end{figure}

The plastic strain, $\varepsilon_p$, is the response function in the DD simulations, and is calculated from the aggregate area swept by each dislocation segment during a time iteration \cite{arsenlis2007enabling,jamond2016consistent}. As such, in the DD/kMC simulations we track $\varepsilon_p$ as a function of time at different temperatures and stresses. A representative example is shown in Figure \ref{straintime}, where we plot the plastic strain versus time at different temperatures for a dislocation source with length $L=150b$ containing two super-jogs under 1500 MPa of applied stress. While the $\varepsilon_p$-$t$ curve for $\tau=500$ K is smooth, indicating the absence of super-jog jumps, the rest of the curves display some roughness associated with sharp super-jog transitions of magnitude $\pm b$. Note that these transitions are both stress and temperature-assisted, as determined by eq.\ \eqref{sinh}.

The graph shows that, for a fixed dislocation length and super-jog separation, higher temperatures lead to a faster dislocation evolution. This can all be attributed to the diffusional part of the dislocation dynamics, as the viscous contribution, governed by $B^{\rm edge}$, displays practically no thermal dependence. 

\subsection{Strain rate-temperature relations}

\begin{figure}[ht!]
   \centering
\subfloat[\label{alana}]{\includegraphics[height=0.4\textwidth]{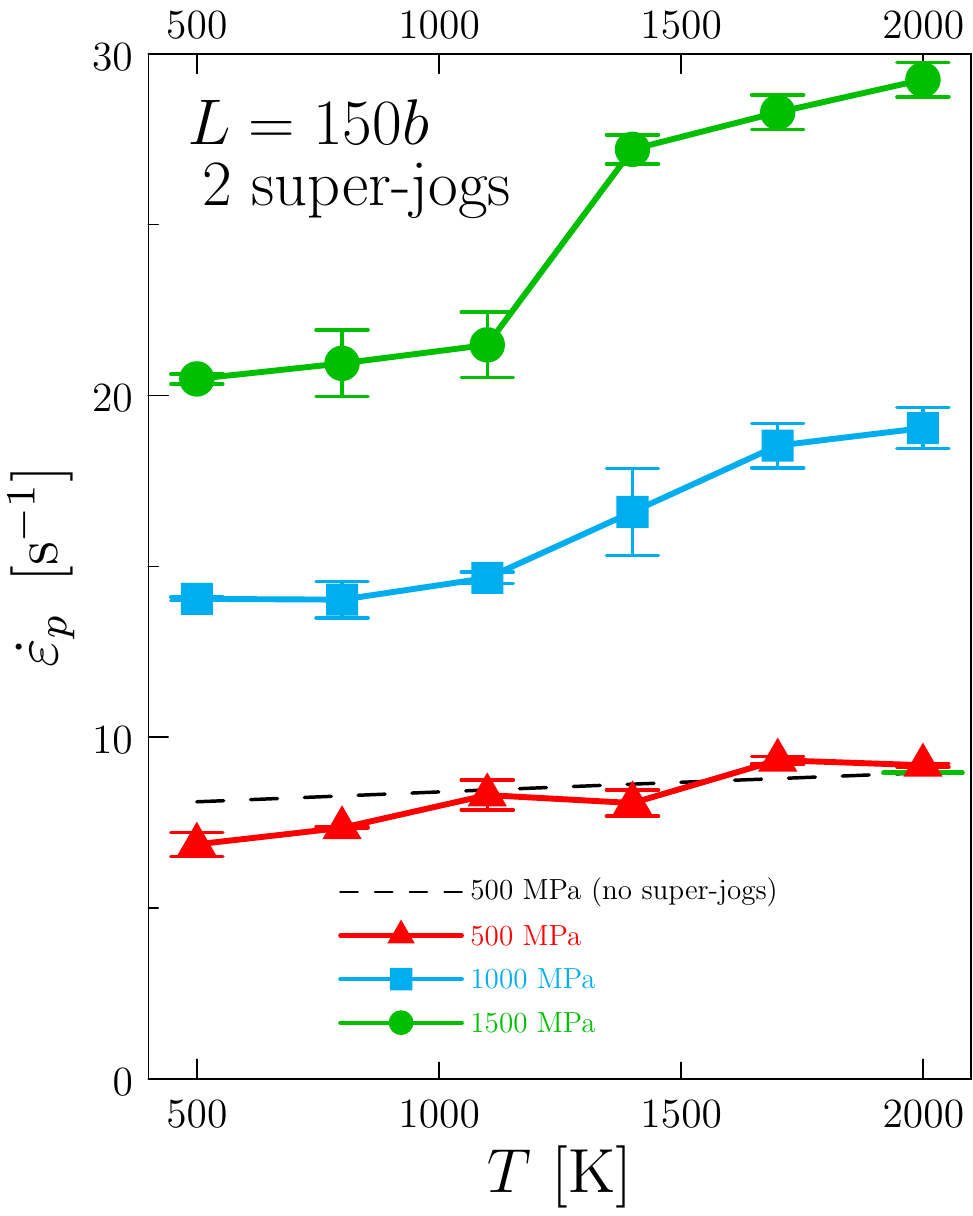}} 
  \subfloat[\label{alanb}]{\includegraphics[height=0.4\textwidth]{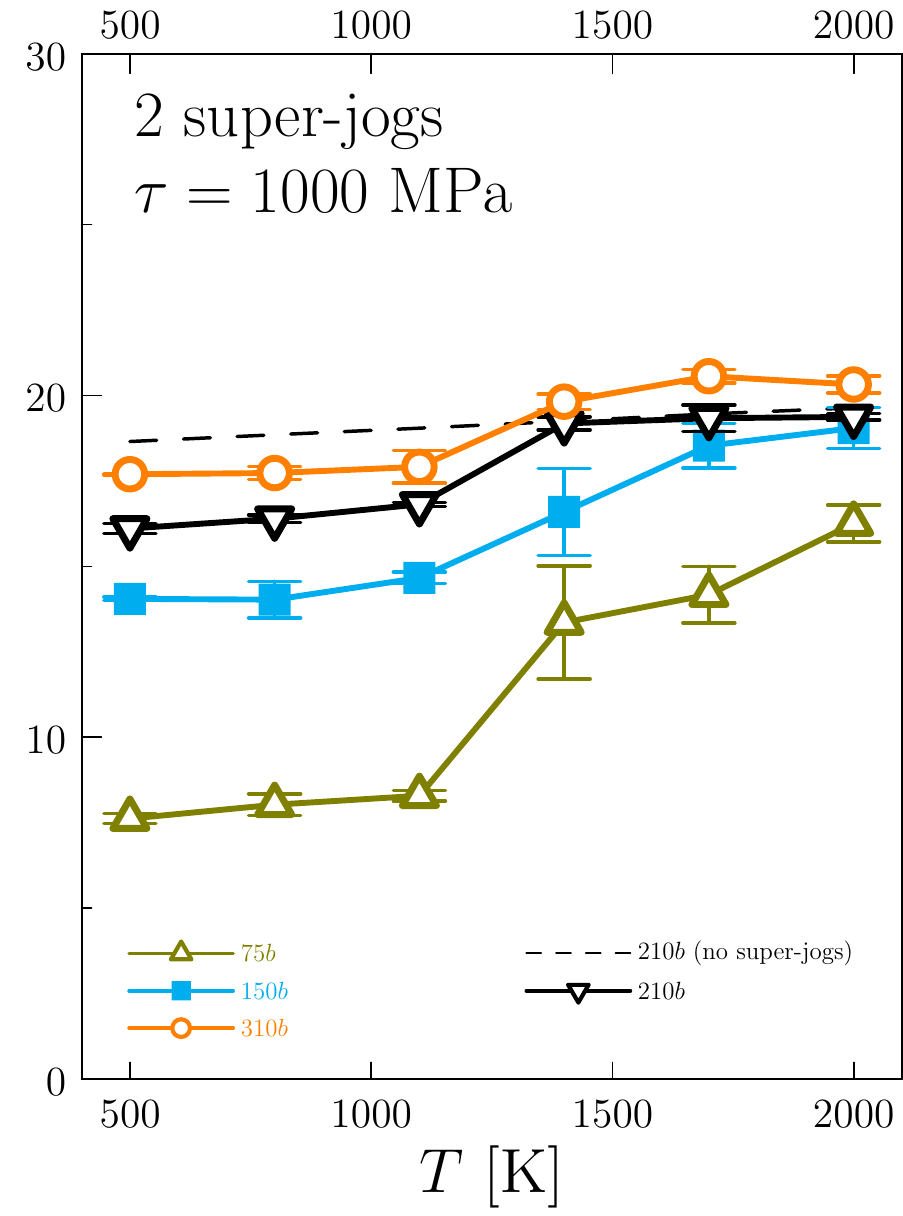}} 
  \subfloat[\label{alanc}]{\includegraphics[height=0.4\textwidth]{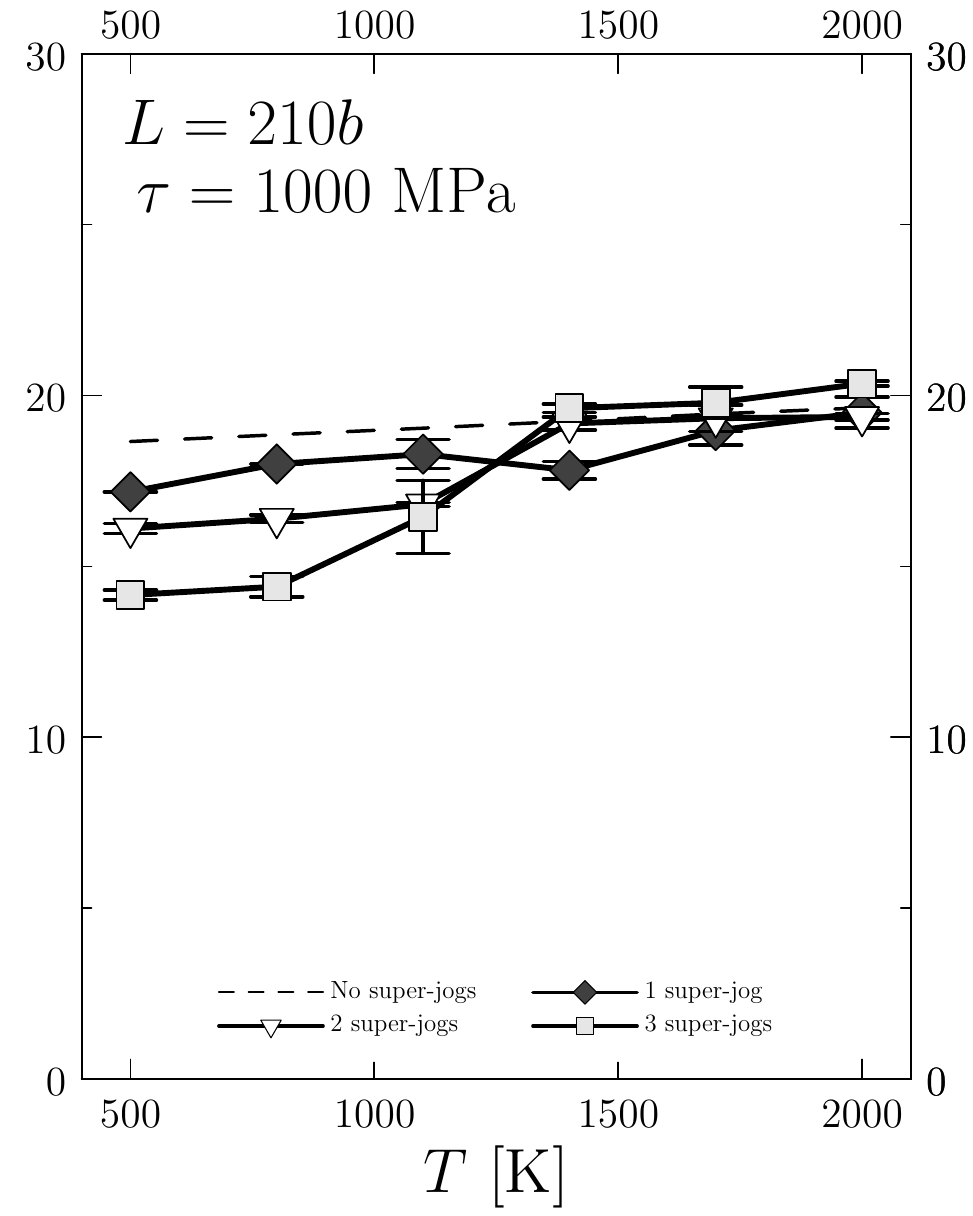}} 
   \caption{Plastic strain as a function of: (a) stress for a fixed line length of $L=150b$ and fixed number of super-jogs (two), (b) line length for a fixed applied stress of 1000 MPa and fixed number of super-jogs (two), and (c) number of super-jogs for fixed stress (1000 MPa) and line length ($210b$). The error bars represent the numerical variability obtained from five independent simulations for each case.}
   \label{alan2}
\end{figure}

First it is useful to note that the cases without super-jogs are trivially modeled by the Orowan equation for the ideal case of infinite dislocation lengths:
\begin{equation}
\dot{\varepsilon}_p=\rho_d b v = \frac{\rho_d b^2}{B^{\rm edge}} \left(\tau-\Delta \tau-\tau_c^{\rm edge}\right)=\frac{\rho_d b^2}{B^{\rm edge}} \left(\tau - \frac{\alpha \mu b}{L}-\tau^{\rm edge}_c\right)
\label{linearstrength}
\end{equation}
where, from Fig.\ \ref{crss}, the only temperature dependence comes via the critical stress $\tau_c$ ($B^{\rm edge}$ is a constant, cf.~Table \ref{tab1}). As such, the (plastic) strain rate given by eq.\ \eqref{linearstrength} is linear both in $\tau$ and $T$ with a proportionality constant independent of temperature. From the values in Tables \ref{tab1} and \ref{param2} and $1/L=\sqrt{\rho_d}$:
\begin{equation}
\dot{\varepsilon}_p=3.72\times10^5(\tau-850+0.1T)~~\left[{\rm s}^{-1}\right]
\label{ratelin}
\end{equation}
Next we analyze all the $\varepsilon_p$-$t$ curves of dislocation sources in the $75<L<310b$ range to extract the relationship between the plastic strain rate, $\dot{\varepsilon}_p$, with temperature and stress. The results are shown in Figure \ref{alan2}. Figure \ref{alana} gives the stress dependence of $\dot{\varepsilon}_p$ for a fixed line length of $150b$ containing two super-jogs. Figure \ref{alanb} shows its dependence with line length (all containing two super-jogs) for a fixed applied stress of 1000 MPa. Finally, Figure \ref{alanc} gives the $\varepsilon_p$-$t$ relation as a function of the number of super-jogs on a $210b$ line at 1000 MPa. 
All three figures show cases of straight dislocations containing no-super-jogs for comparison. As expected, the super-jogs confer an extra strength to the dislocation, clearly seen at lower temperatures. This extra strength is negated at above 1100$\sim$1300 K, when super-jog diffusion is becomes prolific at high  temperatures. Thus, in general, the DD/kMC simulations validate the notion that as the temperature increases the super-jogs lose their strengthening power thanks to stress-assisted diffusion. However, that must be coupled to the fact that, as the temperature increases, dislocation lines contain more thermal super-jogs (see Fig.\ \ref{n0}), which results in higher strength (i.e., lower plastic strain rates).

\section{Discussion}

\subsection{Physical model}

The framework that we use in this paper to study edge dislocation dynamics in RMEA is that of thermally-activated processes modeled using the residence-time algorithm. Thermally-activated events dictate the global timescale of the system, which evolves under elastic forces using the DD method between events. In other words, DD relaxation periods are subordinated to take place within time intervals prescribed by the kMC simulator. This generally fits the timescales on which both methods evolve the dislocation, as typical DD time steps in our calculations are on the order of $10^{-10}$ s, while sampled time steps from kMC are in the $10^{-3}$ to $10^{-8}$-s range between 500 and 1700 K. The only exception is 2000 K, where the event times are on the order of $10^{-12}$ s. There, the dislocation evolution is governed by super-jog jumps, without time for elastic relaxation. 

We have used a SNAP potential for the Nb-Mo-Ta-W system \cite{snap,li2020complex} as the underlying `first-principles' model to calculate defect energetics. This effectively supersedes models that assume that a multielement alloy can be decomposed into a substrate or matrix that possesses the average properties of its elementary constituents and on which every lattice atom constitutes a solute atom \cite{Varvenne:222039,maresca2020mechanistic}. Defect energies do reveal a `high entropy' effect, i.e., they display statistical averages that deviate beyond numerical error from weighted averages of single-valued energies of the individual alloy elements.  Specifically, uniform sampling of the energy distributions results in disproportionate numbers of vacancies with energies below the distribution's mean. What is more, in certain conditions, vacancies appear spontaneously (i.e., athermally) as a consequence of occasional samplings that lead to zero or even negative formation energy values. This effect confers a very particular nature to the Nb-Mo-Ta-W alloy that cannot be surmised from `average' material properties.

What is true for bulk systems is also true for atomistic environments surrounding edge dislocation cores, i.e., vacancies exist thermally (and even spontaneously) on the dislocation line. 
These vacancies on dislocation lines are topologically equivalent to super-jogs, consistent with a number of works in the literature \cite{shiotani1967mechanism,feng1999cross,kitajima2009yielding,srivastava2018unveiling}. It is clear that these super-jogs act as pinning points for the dislocation line, increasing the activation stress and strengthening the material \cite{PhysRev.86.52,SAKA1973,martinez2008,song2017}.
The athermal presence of super-jogs can also be put in the context of the relaxed configurations of the lines. The evidence from numerous MD studies in several RMEA consistently points to a ground state of dislocations characterized by rough line shapes \cite{rao2017atomistic,rao2019modeling,maresca2020theory,chen2020unusual}. Such states are likely to be micro-states, i.e., reflective of the length scale over which compositional fluctuations take place, which is on the order of one atomic distance. Indeed, such roughness is not captured experimentally with conventional microscopy \cite{LILENSTEN2018131,wang2020multiplicity}. From a topological point of view, this roughness may manifest itself as \emph{wiggles} along the dislocation line on the glide plane (i.e., kinks) or as steps on the extra half-plane (jogs). While in our model the effect of the line roughness on the glide plane is subsumed into the temperature dependence of the critical stresses and dislocation friction coefficients, roughness in the form of (super)jogs in the extra half-plane is captured explicitly. This is reminiscent of the existence of cross-kinks in screw dislocation lines in thermal equilibrium recently seen in medium-entropy bcc alloys \cite{zhou2021cross}.

Super-jog segments display two potential modes of motion. One is along the line, requiring \emph{climb} by emission or absorption of vacancies. However, our atomistic calculations show that there is no energy benefit in having a super-jog expand laterally by vacancy emission compared to having two super-jogs in adjacent positions on the line. As such, this mode of motion is no different than allowing for natural (thermal) nucleation of super-jogs along the dislocation line. The second degree of freedom is the one considered in our model, i.e., forward/backward translation of a super-jog by a diffusive process. Such process is strongly influenced by the resolved shear stress, favoring motion in the direction along which it is applied. The migration energies for this mode of motion are conceptually equivalent to vacancy migration energies from a position on the dislocation line to the next lattice site along the glide direction, i.e., one Burgers vector distance, which is of course the atomic jump distance in the bcc lattice. As we will show in the next section, this motion leads to a softening of the increased activation stresses due to super-jog pinning.

\subsection{An analytical model of edge dislocation strength based on the present numerical results}

\begin{figure}[ht]
   \centering
\includegraphics[width=0.75\textwidth]{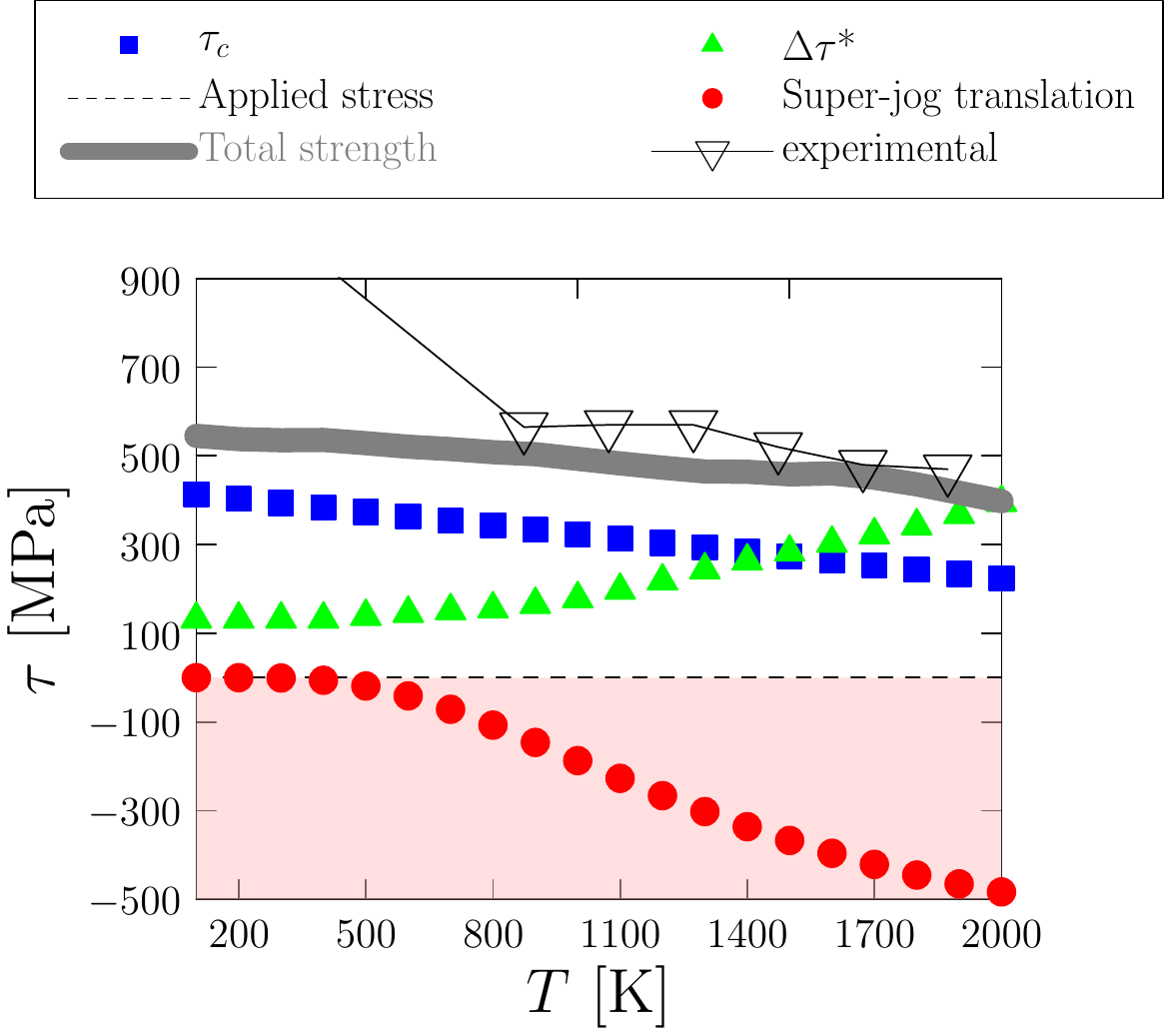}
   \caption{Variation of $\tau$ with temperature together with different terms contributing to it in eq.\ \eqref{stren1} for a strain rate of $\dot{\varepsilon}=10^{-4}$ s$^{-1}$. Experimental data from refs.\ \cite{SENKOV2011698,senkov_miracle_chaput_couzinie_2018}.}
   \label{analytical}
\end{figure}

Using the various contributions to the dislocation evolution model presented in Sec.\ \ref{monte}, a compact expression for the strength of the material due to edge dislocations can be derived (full derivation provided in \ref{app1}): 
\begin{equation}
\tau = \tau_c^{\rm edge}(T) + \Delta\tau^\ast(T) + B^{\rm edge}(T)\left[\frac{\dot{\varepsilon}_0}{\rho_d b^2}-2q^{\rm jog}_m(T)\right]
\label{stren1}
\end{equation}
where $\tau_c^{\rm edge}$ is the critical stress, Table \ref{crss}, $\Delta\tau^\ast$ represents the super-jog hardening, eq.\ \eqref{taylor2}, the term $\left(B^{\rm edge}\dot{\varepsilon}_0/\rho_d b^2\right)$ represents the driving force, eq.\ \eqref{oro1}, and the term $\left(2B^{\rm edge}q^{\rm jog}_m\right)$, eq.\ \eqref{sinh}, represents stress relief due to super-jog motion.
The variation of $\tau$ with temperature using the parameters given in Table \ref{param2} is shown in Figure \ref{analytical}, along with each separate  contributions from the terms in eq.\ \eqref{stren1}. 
The figure clearly shows that the evolution of the strength is controlled by the interplay between the intrinsic hardening due to super-jog nucleation and softening due to super-jog diffusion. This interplay results in a monotonic decrease of $\tau$ with $T$ as shown in the figure. Note that the effect of the applied stress is negligible compared to super-jog-related mechanisms, amounting to no more than 1 MPa at a strain rate of $10^3$ s$^{-1}$.
Most importantly, the strength decreases slowly with temperature, bolstered by a higher thermal concentration of super-jogs (leading to more pinning points) as $T$ increases, and modulated by an easier diffusive glide of the super-jogs with temperature. Such behavior bears a significant resemblance with the temperature dependence of the yield strength of a number of different RMEA \cite{SENKOV2011698,juan2015enhanced,senkov2018development,zhang2021high,feng2021superior}. 
As such, we believe that these mechanisms explain --to a large degree-- the high temperature strength of these alloys, as quantitatively confirmed by our model and calculations. The obtained response is reminiscent of the high temperature behavior of Ni-based superalloys, where the strength results from a balance between the formation and dissolution of Kear-Wilsdorf locks, both of which are temperature-enhanced \cite{vattre2009dislocation}.

It is also interesting to derive an expression for the extra strengthening of a RMEA relative to a pure bcc metal under the same loading conditions (cf.~\ref{app1}):
\begin{equation}
\Delta\tau_{\rm RMEA}(T) \approx \tau_{\rm RMEA}(T)-\tau_{\rm bcc}(T) = \tau_c^{\rm edge}(T) + 
\Delta\tau^\ast(T) -2 B^{\rm edge}(T)q^{\rm jog}_m(T)
\label{stren2}
\end{equation}
This expression gives the terms that contribute to hardening due to the chemical nature of RMEA versus pure or dilute bcc systems. As the equation shows, all terms display a temperature dependence. $\tau_c^{\rm edge}$ displays a weak one in the manner typical of thermal softening, while the other two terms display quasi-exponential (Arrhenius) dependencies. Equation \eqref{stren2} shows that the commonly cited mechanism of lattice fluctuation interactions (local solute interactions) is only one of several terms that contribute to the strength.  We believe that non-conservative processes that are unique to RMEA play a fundamental role in explaining the temperature dependence of the alloy strength.

\subsection{Discussion on model predictions}\label{limit}

Our model results show that the contribution of edge dislocations to the strength of Nb-Mo-Ta-W stems from the balance between two opposing mechanisms. The first is a strengthening contribution associated with a thermal concentration of super-jogs which increases with temperature. The second is a softening effect brought about the thermal motion of the super-jogs, which also increases with temperature. The combination of these two contrary effects dictates the temperature dependence of the edge dislocation contribution. 

In terms of comparison with experimental data, we add the yield strength measurements in equiatomic Nb-Mo-Ta-W alloys in Fig.\ \ref{analytical} \cite{SENKOV2011698,senkov_miracle_chaput_couzinie_2018}. We see that, while below 800 K there is a gap of over 500 MPa between our model and the experimental measurements, at higher temperatures the match is almost exact. Evidently, the alloy yield strength measured experimentally includes other sources of strengthening beyond edge dislocations, particularly at low temperatures where contribution from screw dislocations, small intermetallic particles and/or short range order could play a role. While this may explain the observed gap, it is clear that edge dislocations should be counted as a significant source of strength in this and other similar RMEA.

We end with a note on the role played by the dislocation density in the strength of edge dislocations. As mentioned in Sec.\ \ref{ddd}, values of $\rho_d$ as low as $5\times10^{12}$ m$^{-2}$ have  been measured for Nb-Mo-Ta-W \cite{zou2015ultrastrong}. However, here we have used more realistic values in the vicinity of $10^{15}$ m$^{-2}$ to be consistent with most experimental data in bcc RMEA. In eq.~\eqref{stren1}, the term $\left(B^{\rm edge}\dot{\varepsilon}_0/\rho_d b^2\right)$ represents the driving force and is the only one with an explicit dislocation density dependence. The stress due to this term ranges between $2.7\times10^{-9}$ and $5.4\times10^{-7}$ MPa for $\rho_d=10^{15}$ and $5\times10^{12}$ m$^{-2}$, respectively, when $\dot{\varepsilon}_0=10^{-3}$ s$^{-1}$. This effectively renders the driving force term irrelevant in the global picture of strength, which makes the value of the dislocation density in the range of interest also irrelevant.

\section{Conclusions}
We finish with our main conclusions:
\begin{enumerate}
\item We have proposed a new mechanism for edge dislocation dynamics in Nb-Mo-Ta-W alloys. The mechanism provides a qualitative explanation for the elevated intermediate-temperature strength of refractory multi-element alloys, which is governed by non-conservative  edge dislocation dynamics.

\item Our mechanism is postulated on the favorable thermal existence --sometimes even spontaneous-- of vacancies along edge dislocation lines. These vacancies relax into super-jogs that act as extra pinning points that increase the activation stress. At the same time, these super-jogs are able to diffuse along the glide direction, relieving some of this extra stress. The total strength is a balance between these two processes added on top of the lattice friction due to chemical fluctuations and short-range order. 

\item We have developed a numerical model based on a kinetic Monte Carlo module --which captures thermally activated events (super-jog nucleation and translation)-- and a discrete dislocation dynamics module --which evolves the line configuration in response to elastic forces-- that are coupled to one another via timescale evolution. The global timescale is set by the kMC module on the basis of a Poisson sampling of the thermally activated event rates, while the DD subcycle takes place in between KMC events.

\item All material parameters and alloy energetics have been obtained using a machine-learning SNAP potential. The vacancy formation and migration energies are characterized by distributions with means that deviate from the weighted averages of the elemental constituents of the alloy. Thermal sampling of these distributions results in an equilibrium super-jog concentration that is substantially larger than those predicted by the mean formation energy.

\item An analytical model that captures the essential features of the system yields a strength temperature dependence in excellent qualitative agreement with experimental measurements.
\end{enumerate}

\section*{Acknowledgements}
This work has been funded by the National Science Foundation under Grant No.~DMR-1611342.  We thank UCLA's IDRE for computer time allocations on the \texttt{Hoffman2} supercomputer. 

\section*{References}

\appendix

\section{Computational flow diagram}\label{app2}

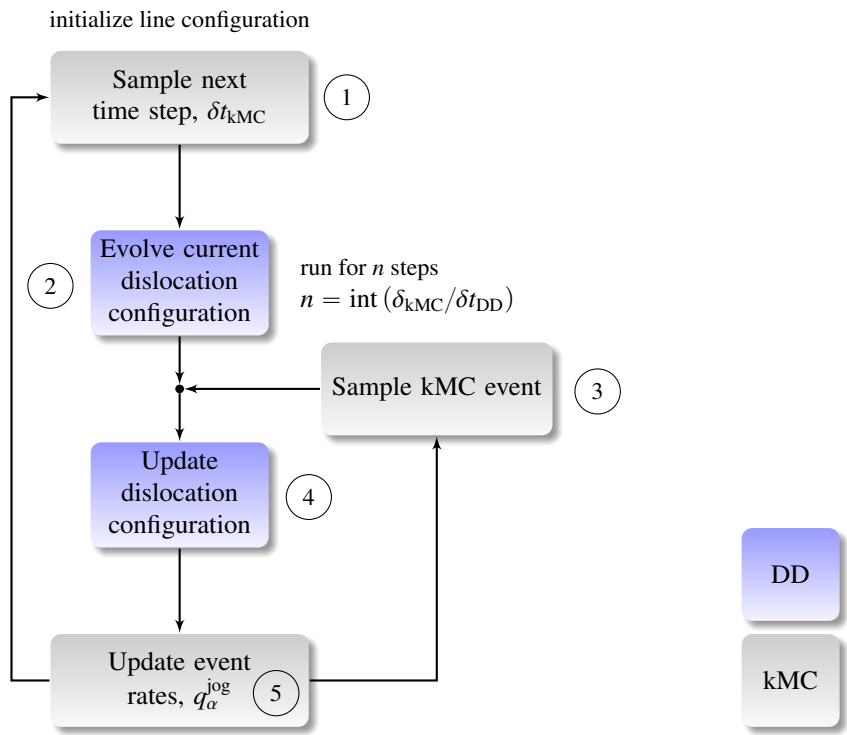
\begin{figure}[h]
\centering
\begin{tikzpicture}[auto]
    \node [box, draw=none, text width=9em, shade, top color=gray!40, bottom color=gray!5,blur shadow={shadow blur steps=5}] (kmc1) {Sample next time step, $\delta t_{\rm kMC}$};
    \node [box, draw=none, below of=kmc1, shade, top color=blue!40, bottom color=blue!5,blur shadow={shadow blur steps=5}] (dd1) {Evolve current dislocation configuration};
    \node [draw, shape = circle, fill = black, below of=dd1, node distance=4em, minimum size = 0.1cm, inner sep=0pt] (c0) {};
        \node [box, draw=none, below of=c0, node distance=4em, shade, top color=blue!40, bottom color=blue!5,blur shadow={shadow blur steps=5}] (dd2) {Update dislocation configuration};
    \node [box, draw=none, text width=9em, below of=dd2, shade, top color=gray!40, bottom color=gray!5,blur shadow={shadow blur steps=5}] (kmc3) {Update event rates, $q^{\rm jog}_\alpha$};
 
    \coordinate (middleup) at ($(kmc1.west)!0.5!(dd1.west)$);
    \coordinate (middle) at ($(dd1.west)!0.5!(dd2.west)$);
    \coordinate (down) at ($(kmc3.east)$);
     \node [box, draw=none, right of=middle, text width=8em, node distance=13em, shade, top color=gray!40, bottom color=gray!5,blur shadow={shadow blur steps=5}] (kmc2) {Sample kMC event};


    \path [line] (kmc1) -- (dd1);
    \path [line] (dd1) -- (c0);
    \path [line] (c0) -- (dd2);
    \path [line] (dd2) -- (kmc3);
   \path [line] (kmc2) -- (c0);
\draw[line]    (kmc3.west) -- + (-0.5,0) |- (kmc1);

    \path [line] (kmc3.east) -| (kmc2.south);
\node[color=black] at (0,1) {\small initialize line configuration};
\node[color=black,align=left] at (3.0,-2.5) {\small run for $n$ steps\\$n={\rm int}\left( \delta_{\rm kMC}/\delta t_{\rm DD}\right)$};

\node[draw,circle] (step1) at (2.2,0)  {\small 1};
\node[draw,circle] (step2) at (-1.7,-2.5)  {\small 2};
\node[draw,circle] (step3) at (5.5,-3.9)  {\small 3};
\node[draw,circle] (step4) at (1.7,-5.3)  {\small 4};
\node[draw,circle] (step5) at (1.27,-7.9)  {\small 5};

     \node [box, draw=none, right of=down, text width=3em, node distance=18em, shade, top color=gray!40, bottom color=gray!5,blur shadow={shadow blur steps=5}] (kmc4) {kMC};
\node [box, draw=none, above of= kmc4, text width=3em, node distance=4em, shade, top color=blue!40, bottom color=blue!5,blur shadow={shadow blur steps=5}] (dd4) {DD};

\end{tikzpicture}
\caption{Flow diagram of the numerical procedure employed here. Processes pertaining to the kMC module are colored in shaded gray, while those pertaining to the DD module are colored in shaded blue. Each box is numbered according to the sequence of steps.\label{diagram}}
\end{figure}

\section{Analytical model of edge dislocation strength due to super-jog evolution}\label{app1}

For an edge dislocation source with a length $\ell$, one can calculate the activation stress in the standard way as:
$$\Delta \tau = \frac{\alpha\mu b}{\ell}$$
which is independent of temperature if we neglect the thermal softening of the elastic constants of the material. When super-jogs appear, the total dislocation length is shortened into segments of average length $\ell^\ast$:
\begin{equation}
\Delta \tau^\ast(T) = \frac{\alpha\mu b}{\ell^\ast(T)}
\label{taylor1}
\end{equation}
where $\ell^\ast(T)$ is the inverse of the super-jog concentration at temperature $T$:
\begin{equation}
\ell^\ast(T)=n_0(T)^{-1}=w\exp\left(\frac{\bar{E}^\perp_f}{kT}\right)
\label{arrhen}
\end{equation}
where we have used $n_0^\ast=\ell/w$ as the pre-factor of $n_0(T)$, which is equal to the total number of independent nucleation sites along the dislocation line. Equation \eqref{taylor1} then becomes:
\begin{equation}
\Delta \tau^\ast(T) = \frac{\alpha\mu b}{w}\exp\left(-\frac{\bar{E}^\perp_f}{kT}\right)
\label{taylor2}
\end{equation}
where we are assuming an exponential temperature dependence of $n_0(T)$ defined by the mean energy value of $p(E^\perp_f)$ (Fig.\ \ref{edpos}). Below, this will be replaced with the actual dependence of $n_0$ on $T$ obtained in Fig. \ref{n0}.
The \emph{excess} stress available for dislocation glide is then obtained by subtracting $\Delta \tau^\ast(T)$ and $\tau^{\rm edge}_c(T)$ (Table \ref{tab:param}) from the resolved shear stress $\tau$:
\begin{equation}
\Delta\tau_{\rm gl} = \tau - \left(\Delta \tau^\ast(T) + \tau^{\rm edge}_c(T)\right)
\label{taylor3}
\end{equation}
The glide velocity can then be defined as:
\begin{equation}
v_{\rm gl}=\frac{b\Delta\tau_{\rm gl}}{B^{\rm edge}(T)}
\label{vglide}
\end{equation}
Concurrently, super-jogs can advance along the glide direction with a velocity given by:
\begin{equation}
v_{\rm sj}=bq^{\rm jog}_m=2b\nu_{0}'\exp\left(-\frac{\bar{E}^\perp_m}{kT}\right)\sinh\left(\frac{1.2\Omega_a\tau}{kT}\right)
\label{vsj}
\end{equation}
where the `$\sinh$' term captures the balance of forward and backward jumps. Here again, we take $\bar{E}^\perp_m$ as representative of the $p(E^\perp_m)$ energy distribution (Fig.\ \ref{edform}).
The total velocity of the dislocation line is:
\begin{equation}
v_{\rm tot}=v_{\rm gl}+v_{\rm sj}
\label{vtot}
\end{equation}
which can be related to a prescribed strain rate $\dot{\varepsilon}_0$ using Orowan's equation:
 \begin{equation}
\dot{\varepsilon}_0=b\rho_d v_{\rm tot}
\label{oro1}
\end{equation}
where $\rho_d$ is the dislocation density. Combining eqs.\ \eqref{vglide}, \eqref{vsj}, and \eqref{oro1}, we arrive at:
 \begin{equation}
\dot{\varepsilon}_0=b\rho_d\left[\frac{b\Delta\tau_{\rm gl}}{B^{\rm edge}(T)}+2b\nu_{0}'\exp\left(-\frac{\bar{E}^\perp_m}{kT}\right)\right]
\label{oro2}
\end{equation}
Operating and using the various definitions for the different terms in the above equations, we can write a compact expression for the strength of the material due to edge dislocations: 
\begin{equation}
\tau = \tau_c^{\rm edge}(T) + \Delta\tau^\ast(T) + B^{\rm edge}(T)\left[\frac{v_{\rm tot}}{b}-2q^{\rm jog}_m(T)\right]
\label{strengtheq}
\end{equation}
with:
\begin{eqnarray*}
\tau_c^{\rm edge}(T)&=&423.6 - 0.1T ~{\rm [MPa]}\\
\Delta\tau^\ast(T) &=& \frac{\alpha \mu b}{w}\exp\left(-\frac{\bar{E}^\perp_f}{kT}\right)\\
v_{\rm tot} &=& \frac{\dot{\varepsilon}_0}{\rho_d b}\\
q^{\rm jog}_m(T)&=&\nu_{0}'\exp\left(-\frac{\bar{E}^\perp_m}{kT}\right)\sinh\left(\frac{1.2\Omega_a\tau}{kT}\right)
\end{eqnarray*}
For convenience, we list all the material constants in Table \ref{param2}.
\begin{table}[ht!]
\caption{Parameters and material constants used to evaluate eq.~\eqref{strengtheq}. \label{param2}}
\begin{center}
\begin{tabular}{|c|c|c|c|c|}
\hline
Parameter & Description & Value & Units & Source \\
\hline\hline
$\alpha$ & hardening coefficient & 0.5 & -- & this work \\
$\mu$ & shear modulus & 94 & GPa & \cite{li2020complex} \\
$a_0$ & lattice parameter & 3.24 & \AA & this work \\
$b$ & $a_0\sqrt{3}/2$ & 2.81 & \AA & this work \\
$w$ & $a_0\sqrt{6}/3$ & 2.64 & \AA & this work \\
$\Omega_a$ & $a_0^3/2$ & $1.7\times10^{-29}$ & m$^3$ & this work \\
$B^{\rm edge}$ & dislocation friction coefficient & $2.12\times10^{-4}$ & Pa$\cdot$s & \cite{yin2021atomistic} \\
$\bar{E}^\perp_f$ & effective super-jog formation energy & 0.02$\sim$0.13 & eV & this work (Fig. \ref{n0}) \\
$\rho_d$ & dislocation density & $1\sim2\times10^{15}$ & m$^{-2}$ & Sec.~\ref{ddd} \\
$\nu_0'$ & attempt frequency & $10^{13}$ & Hz & this work \\
$\bar{E}^\perp_m$ & effective super-jog migration energy & 0.32 & eV & this work (Fig. \ref{edform}) \\
\hline
\end{tabular}
\end{center}
\label{default}
\end{table}%
Equation \eqref{strengtheq} shows that the sources of strengthening for edge dislocations are the intrinsic lattice stress, $\tau_c^{\rm edge}$ (decreases linearly with temperature), the extra stress due to the existence of super-jogs, $\Delta\tau^\ast(T)$ (increases exponentially with temperature), and the applied stress, $\dot{\varepsilon}_0B^{\rm edge}/\rho_d b^2$ (independent of temperature). Conversely, the motion of super-jogs, represented by $q^{\rm jog}_m$, reduces the strength of the material (increasing also exponentially with temperature). However, to obtain the true temperature dependence of the strength of the system, we must substitute the standard Arrhenius expression for $n_0(T)$ in $\Delta\tau^\ast(T)$, eqs.\ \eqref{arrhen} and \eqref{taylor2}, with its actual thermal dependence given in Fig.\ \ref{n0}. Such substitution gives rise to the temperature response shown in Fig.\ \ref{analytical}. 

It is also interesting to derive the extra strengthening due to multielement alloy effects. The equivalent of eq.\ \eqref{strengtheq} for a pure system or dilute alloy would read:
\begin{equation}
\tau_{\rm bcc} = \tau_c^{\rm edge}(T) + \alpha\mu b\sqrt{\rho_d} + B^{\rm edge}(T)\frac{\dot{\varepsilon}_0}{b^2\rho_d}
\label{strengtheq1}
\end{equation}
i.e., $\Delta\tau(T)^\ast\equiv\Delta\tau(T)$ takes the standard Taylor form ($\alpha\mu b\sqrt{\rho_d}$) and becomes independent of temperature. Assuming, as it is customary, that $\tau_c^{\rm edge}\approx 0$ in bcc metals, and that $w\ll\rho_d^{-1/2}$, i.e., neglecting the Taylor hardening term, we can subtract eq.\ \eqref{strengtheq1} from \eqref{strengtheq}, we arrive at the extra strengthening associated with RMEA effects:
\begin{equation}
\Delta\tau_{\rm RMEA}(T) \approx \tau_{\rm RMEA}(T)-\tau_{\rm bcc}(T) = \tau_c^{\rm edge}(T) + 
\frac{\alpha\mu b}{w}\exp\left(-\frac{\bar{E}^\perp_f}{kT}\right)-2B^{\rm edge}\nu_{0}'\exp\left(-\frac{\bar{E}^\perp_m}{kT}\right)\sinh\left(\frac{1.2\Omega_a\tau}{kT}\right)
\label{strengtheq2}
\end{equation}
This expression gives the terms that contribute to hardening due to the chemical nature of RMEA versus pure or dilute bcc systems. As the equation shows, only the last term has an explicit dependence on stress. In other words, the difference between RMEA and bcc systems decreases with increasing stress on account of the super-jog diffusion term.

\end{document}